\UseRawInputEncoding
\documentclass[journal]{IEEEtran}
\usepackage{amsmath,amsfonts}
\usepackage{algorithmic}
\usepackage{algorithm}
\usepackage{array}
\usepackage[caption=false,font=normalsize,labelfont=sf,textfont=sf]{subfig}
\usepackage{textcomp}
\usepackage{stfloats}
\usepackage{url}
\usepackage{verbatim}
\usepackage{graphicx}
\usepackage{cite}
\usepackage{graphicx}
\usepackage{subfig}
\usepackage{multicol}
\usepackage{changepage}
\usepackage[intoc, english]{nomencl}
\usepackage{blindtext}
\usepackage{setspace}
\usepackage{empheq}
\usepackage{optidef}
\usepackage{xcolor}
\usepackage{amsmath,amsfonts,amssymb}
\usepackage{amsthm}
\usepackage{mathtools}
\usepackage{nomencl}
\usepackage{etoolbox}
\usepackage{hyperref}
\renewcommand\nomgroup[1]{%
  \item[\bfseries
  \ifstrequal{#1}{P}{Superscript}{%
  \ifstrequal{#1}{B}{Subscripts}{%
  \ifstrequal{#1}{O}{Other symbols}{}}}%
]}
\newtheorem{remark}{Remark}
\newtheorem{prop}{Proposition}
\newtheorem{property}{Property}

\begin{document}   
\title{Perch like a bird: bio-inspired optimal maneuvers and nonlinear control for Flapping-Wing Unmanned Aerial Vehicles} 
\author{C. Ruiz and J. \'{A}. Acosta
\thanks{Dept. de Ingenier\'ia de Sistemas y Autom\'atica, Universidad de Sevilla (Spain). Corresponding author: crpaez@us.es (C. Ruiz)}
\thanks{This work has been supported by the emerging research group Multi-robot And Control Systems (\href{https://prisma.us.es/colectivo/grupo/TEP-995}{MACS}) under the VII PPIT-US and  VI PPIT-US framework, University of Seville (SPAIN).

The authors thank Dr. M. Klein H. B. from the Department of Zoology, University of Oxford, for their valuable discussion on \cite{hawk_perching}.}
}
\markboth{\parbox{0.99\textwidth}{\textit{C. P\'aez \& J.\'A. Acosta} -- This work has been submitted to the IEEE for possible publication. Copyright may be transferred without notice, after which this version may no longer be accessible.}}{}

\maketitle

\begin{abstract}
This research endeavors to design the perching maneuver and control in ornithopter robots. By analyzing the dynamic interplay between the robot's flight dynamics, feedback loops, and the environmental constraints, we aim to advance our understanding of the perching maneuver, drawing parallels to biological systems.
Inspired by the elegant control strategies observed in avian flight, we develop an optimal maneuver and a corresponding controller to achieve stable perching. The maneuver consists of a deceleration and a rapid pitch-up (vertical turn), which arises from analytically solving the optimization problem of minimal velocity at perch, subject to kinematic and dynamic constraints. The controller for the flapping frequency and tail symmetric deflection is nonlinear and adaptive, ensuring robustly stable perching. Indeed, such adaptive behavior in a sense incorporates homeostatic principles of cybernetics into the control system, enhancing the robot's ability to adapt to unexpected disturbances and maintain a stable posture during the perching maneuver.
The resulting autonomous perching maneuvers---closed-loop descent and turn---, have been verified and validated, demonstrating excellent agreement with real bird perching trajectories reported in the literature.
These findings lay the theoretical groundwork for the development of future prototypes that better imitate the skillful perching maneuvers of birds.
\end{abstract}

\section{Introduction}\label{sec:intro}
\IEEEPARstart{C}{ybernetics}, the study of communication and control in living organisms and machines, provides a powerful framework for understanding the complex challenges facing controlled maneuver in ornithopters, i.e. bio-inspired aerial vehicles that mimic the flight of birds. 
However, in order to take advantage of ornithopters when compared to other types of Unmanned Aerial Vehicles (UAVs)---like sustainability, safety, efficiency and low environmental impact---, they must be able to perform a controlled landing, so-called perching, and this is the \emph{raison d'\^etre} for the present work.

The most prevalent bioinspired method of landing is perching, which remains an unsolved challenge in robotics due to the high level of control authority required to reduce velocity while maneuvering immersed in complex and unstable flight dynamics within dynamic and kinematic constraints \cite{mechanics_perching}. Currently, prototypes are being developed to perform perching. However, the literature is scarce regarding the bio-informed approach that encompasses both the perching maneuver and its efficient control \cite{bio_informed}.

In this proposal, we aim to bridge the gap between biologists' extensive study of bird flight and control and robotics specialists' development of flight controllers, from a cybernetic perspective.
In addition to the controller's development, bio-inspired maneuvers are analyzed, and an optimal trajectory is first-time proposed for evaluation of control performance. Furthermore, these maneuvers have been validated with real bird flight data from published literature \cite{hawk_perching}. It is noteworthy that wing complex morphing plays a crucial role in perching maneuvers, particularly in enhancing thrust \cite{harvey_morfing,Ruiz2022}. However, to propose an explicit solution to this intricate problem, we restrict ourselves to flexible wings.

%
Bird flight phases are the result of natural adaptation over thousands of years, optimizing a different objective in both morphological and kinetic aspects, 
prompting the pursuit of its discovery \cite{optimal_kinematics}.
The phases of perching vary in number, depending on the author. By image reconstruction, authors in \cite{mechanics_perching} divide the phases into a stationary flight, a highly unsteady pitch-up maneuver and a deep stall. 
In \cite{hawk_perching}, authors analyzed 1576 the perching maneuvers of Harris Hawks birds captured using a Motion Capture System (MCS). They found that large birds' trajectories are divided into two phases: a powered descent and a rapid pitch up. They then created a simplified model with constant lift and power and optimized three functions: energy, time, and distance flown before stalling. Neither time nor energy minimization directly explains the birds' deep swooping flight behavior.
%
\\
Regarding robots perching trajectories have been studied for fixed wing UAVs, using optimization algorithms \cite{optperchfixed} or reinforcement learning \cite{perching_reinforcement}. On the other hand, different clamping methods in development include magnetic contact, pinching, or gripping, so minimal end velocity is desirable to avoid damage. These restrictions, along with flight restrictions, must be considered during maneuvers \cite{Feliu2021}.

\begin{figure*}[ht]
    \centering
\includegraphics[width=0.85\textwidth]{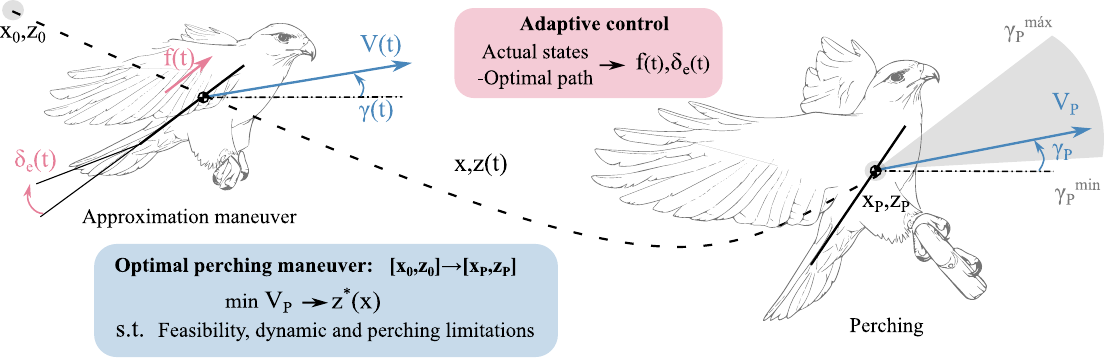}
    \caption{Sketch of the proposed methodology for the autonomous approach for the perching maneuver.}
    \label{fig:pajaro2}
\end{figure*}

In the field of flapping wing flight control for ornithopters, while numerous works exist for cruise flight, the availability of proposals or applications for perching flight maneuvers is relatively limited.
In this work, we opt for a nonlinear adaptive control strategy, because it parameterizes the uncertainty of the model and uses feedback to learn and guarantee the objective \cite{khalil}. Thus, the following requirements are covered, which are partially or non addressed in the literature: {\bf R0)} bird-size Flapping-Wing Unmanned Aerial Vehicles (FW-UAVs) with all processors, sensors and actuators necessary for autonomous outdoor flight on board; {\bf R1)} explicit flight controller where flapping frequency is an input; {\bf R2)} dimensionless model accounting for high angles of attack at low velocities and stall in cruise and perching maneuvers regimes, and tractable for model-based control; {\bf R3)} optimal bio-inspired safe perching trajectories with feasibility guarantees; and {\bf R4)} underactuated FW-UAV. Let us briefly review the literature across the requierements.

The adaptive control technique has been applied to fixed wing UAVs in  \cite{gavilan2015}. Additionally, in a preliminary work \cite{maldonado2020} authors implemented it in the longitudinal gliding dynamics, whose outer controller closed the loop with an indoor MCS, thereby enabling validation with a precision of centimeters. In both, there is no flapping and model for cruise only in the former, thus they do not cover {\bf R1-3} and partially {\bf R0} and {\bf R4}.

In the case of insect-like prototypes the unsteady aerodynamics can be neglected and assumed quasi-steady, hence {\bf R0} is not covered. Thus, e.g. in \cite{adaptive_insect} an adaptive controller was developed with sliding mode in order to follow a path, 
 in \cite{weiNN,chenNN} an adaptive Neural Network controller 
only for cruise flight of micro FW, hence these does not cover {\bf R1-3} either.\\ 
It is worth mentioning the work on developing adaptive controllers to control only the flexible wings mechanism, and therefore outside of any requirement, as in \cite{weiADAPT,weiIT}.


Before presenting the conclusions, we make the following considerations regarding the requirements.
For {\bf R3} an analytical study has been conducted on bio-inspired maneuvers to minimize perching velocity. Realistic perching trajectory optimization faces challenges, including dynamic limitations like maximum deceleration rate and minimum turn radius, and kinematic limitations like stall velocity and perching final pose due to claws. Optimized trajectories closely resemble real bird maneuvers \cite{hawk_perching}. These trajectories resembling birds do not coincide with the attainable ones obtained by trial and error as a preliminary approach, indoors with a MCS and PID feedback in \cite{perching} by some of the authors, which do not cover {\bf R0-2} either.
For {\bf R2} we start with the experimental validated model developed by the authors in \cite{Ruiz2021,guzman2021} and add complex aerodynamic phenomena like stall and flapping interference, valid for any bird-size ornithopter. 
%
On the other hand, {\bf R4} imposes a hard constraint to alleviate weight by non-redundancy of actuators and meet {\bf R0}, at the expense of adding complexity to the controller.

The outline of the proposal is shown in Fig. \ref{fig:pajaro2}, and the thechnical contributions are enumerated in detail below: 

\begin{itemize}
\item[C1.] The optimal perching maneuver for large ornithopters (see Fig. \ref{fig:pajaro2}) based on minimal velocity that accomplish the mechanical, dynamical and kinematic constraints. The dynamical constraints considered allow to obtain a feasibility criteria of the perching, which are scarce in literature and provide a practical rule-of-thumb for their extrapolation to other FW-UAVs.
To the best of the authors' knowledge, the problem has been solved analytically for the first time and the solution performs as the natural perching flights of birds reported  \cite{hawk_perching}.

\item[ C2.] The nonlinear and adaptive control scheme for the FW-UAV and tail-driven underactuated platform. The controller is model free adapting the aerodynamics, and nonlinear, operating even in post-stall regime. 
The scheme has been verified in critical perching maneuvers. For the sake of comparison, the major theoretical control differences between fixed and flapping wing have been highlighted, as the non-autonomous flapping dynamics which enforce a robust performance.

\item[C3.] The unsteady aerodynamic model, which accounts for the nonlinear effects as wing/tail stall at low velocities or the effect of flapping perturbations, while maintaining comprehensiveness, 
thus allowing its use in model-based control design, unlike its counterparts. The full model 
is provided in the document for the community. 
\end{itemize}
The paper is structured as follows: In section \ref{sec:dynmodel} the nonlinear dynamic model of the platform is described. In sec. \ref{Optimal perching maneuvers} we analyze a feasible optimal approach maneuver for perching, including analytical solution verified with real perching hawk data. Section \ref{controller} is devoted to the development of the nonlinear adaptive flight controller, highlighting the requirements accomplishment.
Finally, in sec. \ref{sec:Results} we implement and verify the controller through the optimal trajectory and validate it with real hawk data. Conclusions are summarized in \ref{Conclusions and future work}.

\smallskip
\noindent{\bf Notation.} All vectors are columns. The absolute value of $\xi\in\mathbb{R}$ is denoted as $|\xi|$. The Lebesgue spaces of functions are denoted as $\mathcal{L}_p$ with $p=\{\infty,2\}$. A continuous function $\nu:[0,a)\mapsto [0, \infty)$ belongs to class $\mathcal{K}$, if it is strictly increasing and $\nu(0)=0$. It belongs to class $\mathcal{K}_{\infty}$ if $a=\infty$ and $\nu(r)\to \infty$ as $r\to \infty$.
The symbol $\mathcal{O}(\cdot)$ stands for the order of magnitude such that $\varpi=\mathcal{O}(\epsilon) \Leftrightarrow |\varpi|\leq \kappa|\epsilon|$ for $|\epsilon|\leq\epsilon_{0}$ and $\kappa,\epsilon_{0}>0$.

\nomenclature[P]{\(w,t,b\)}{Subscripts, related to wing, tail and body.}
\nomenclature[N]{\(s\)}{Subscripts, related to aerodynamic stall.}
\nomenclature[N]{\(0,A,P\)}{Subscripts, related to initial, transition and perching points of approach maneuver.}
\nomenclature[N]{\(D,T\)}{Subscripts, related to descent and turning phases of approach maneuver.}
\nomenclature[N]{\(R\)}{Subscripts, reference value.}

\section{Ornithopter dynamics}\label{sec:dynmodel}
In this section, we present the extrapolation to a flapping wing of the well known flight dynamic model. 
The reference platform is the e-flap ornithopter (see \cite{Zufferey2021}), a bird-size FW-UAV. However the methodology presented in this work could be implemented in any bird-size ornithopter, since the aerodynamic model is nondimensional. The main platform parameters are the distance from aerodynamic center (1/4 c) to gravity center, ($x_a$), and from it to tail hinge point ($x_t$), wing and tail chords and surface and wingspan ($c, c_t, S, S_t, b$), the ornithopter inertia and mass ($I_{y}, m$) and the gravity and air density $g$ and $\rho$. For the perching maneuver, we focus on longitudinal dynamics, which are the most relevant. Thus, the minimal set of variables are the states $\mathbf{X} = (x, z, \theta, V, \gamma, q) \in  \mathbb{R}^ 6$ and control inputs $\mathbf{U}=(f, \delta_e) \in  \mathbb{R}^ 2$. The ornithopter longitudinal time-dependent dynamics can be expressed as $
\dot{\mathbf{X}}=\mathbf{H}(\mathbf{X},\mathbf{U},t),
$
where the following definitions are in order:
\begin{itemize}
\setlength\itemsep{-0.2em}
\item[-] $x,z$: Inertial forward/vertical position [$m$].
\item[-] $\theta$: Pitch (positive nose up) [$rad$].
\item[-] $\gamma$: Path angle [$rad$].
\item[-] $V$: Body velocity [$m/s$].
\item[-] $q$: Angular velocity (positive nose up) [$rad/s$].
\item[-] $f$: Flapping frequency [$Hz$].
\item[-] $\delta_e$: Tail deflection, positive tail down [$rad$].
\end{itemize}
Other derived states used are the linear body velocities $u=V\cos{\gamma}$ and $w=V\sin{\gamma}$ (note that $V=\sqrt{u^2+w^2}$), wing and tail angle of attack $\alpha=\arctan(w/u)$, $\alpha_t=\alpha + \frac{x_t q}{V}$, and the reduced frequency  $k=\frac{\pi f c}{V}$. The set of longitudinal equations of motion derived from Newton-Euler become
\begin{empheq}[left=\text{$\Sigma_{p}:$}\empheqlbrace]{align}
&\dot{x} = V \cos{\gamma}, \label{eom1}\\
&\dot{z} = V \sin{\gamma}, \label{eom2}
\end{empheq}
\begin{empheq}[left=\text{$\Sigma_{V}:$}\empheqlbrace]{align}
&\dot{V}= (F_X-mg\sin{\gamma})/m, \label{eom3}
\end{empheq}
\begin{empheq}[left=\text{$\Sigma_{\gamma}:$}\empheqlbrace]{align}
&\dot{\gamma}= (F_Z-mg\cos{\gamma})/(mV), \label{eom4}\\
&\dot{\theta}=q, \label{eom5}\\
&\dot{q}= {F_M}/{I_{y}}, \label{eom6}
\end{empheq} 
where $F_X$, $F_Z$, $F_M$, are the aerodynamic forces
, which depend on the lift, drag and pitch moment, namely $L$, $D$ and $M$, for the wing, tail and body (subscripts $w,t$ and $b$) as follows
\begin{align}
F_X &= - D_w  + L_t \sin{(\alpha_t-\alpha)} 
- D_t \cos{(\alpha_t-\alpha)}  - D_b, \label{F1} \\
F_Z &= L_w + L_t \cos{(\alpha_t-\alpha)} 
+ D_t \sin{(\alpha_t-\alpha)} + L_b, \label{F2} \\
F_M &= L_w x_a \cos{\alpha} + D_w x_a \sin{\alpha} + M_{w} \nonumber \\
&- L_t x_t \cos{\alpha_t}  - D_t x_t \sin{\alpha_t} + M_{t}. \label{F3}
\end{align}
The aerodynamic modeling of these forces have been identified from previous CFD simulations \cite{Ruiz2021,guzman2021}. Regarding wing,
\begin{align*}
    L_w (t) &= {L}_w+\check{L}_w \sin{2 \pi f t}, \\
{L}_w &=q \Theta_{{L}_w}^T r_w(\alpha,k) \mu(\alpha),\\
\check{L}_w &=q \Theta_{\check{L}_w}^T r_w(\alpha,k) \mu(\alpha).
\end{align*}
where $q=\frac{1}{2} \rho V^2 S$, $\Theta_{{L}_w}$ and $\Theta_{\check{L}_w}$ are parameter vector  
and $r_w=(1,\alpha, k ,\alpha^2,k^2,\alpha k,k^3)^T$. Finally, $\mu(\alpha)=\cos^3{(\Theta_S^1 \alpha+\Theta_S^2)}$ is a function modeling the stall effect.
\begin{remark}
It is worth noting that, for the sake of accuracy the wing lift has been modeled as a time-dependent sinusoidal. On the other hand, the remaining loads are considered time invariant since the flapping perturbation is irrelevant when compared to wing lift. This makes the controller design much more involved, being necessary the use of the theory of averaging to overcome the periodic dependence with time (see Section~\ref{controller}).
\end{remark}
The drag and the pitch moment at 1/4c are modeled as:
\begin{align*}
D_w&=q \Theta_{{D}_w}^T r_w(\alpha,k), \\
M_{w}&=qc \Theta_{{M}_{w}}^T r_w(\alpha,k). 
\end{align*}
Note that 
the drag diverges as k increases, so we limit the model to $k<2$. 
 Regarding tail, sinusoidal model has been identified to account for stall, which is critical for control,
\begin{align*}
L_t&=q_t \Theta_{{L}_t}^1 \sin{\big(\Theta_{{L}_t}^2 (\alpha_t + \delta_e )\big)},\\
D_t&=q_t \big [\Theta_{{D}_t}^0- \Theta_{{D}_t}^1 \cos{\big(\Theta_{{D}_t}^2 (\alpha_t + \delta_e )\big)}\big],\\
M_{t}&=q_t c_t \Theta_{{M}_{t}}^1 \sin{\big(\Theta_{{M}_{t}}^2 (\alpha_t + \delta_e)\big)}.
\end{align*}
where $q_t=\frac{1}{2} \rho V^2 S_t$. Note that while wing's model parameters are vectors, the tail aerodynamic model are scalars. The values of the aerodynamic parameters of the complete model are presented in Appendix \ref{ap:aerodynamics}. The body loads are negligible with respect other aerodynamic forces $\mathcal{O}(10^{-1})$.

\section{Optimal perching maneuver}\label{Optimal perching maneuvers}
\begin{figure}[htbp]
    \centering
\includegraphics[width=0.95\columnwidth]{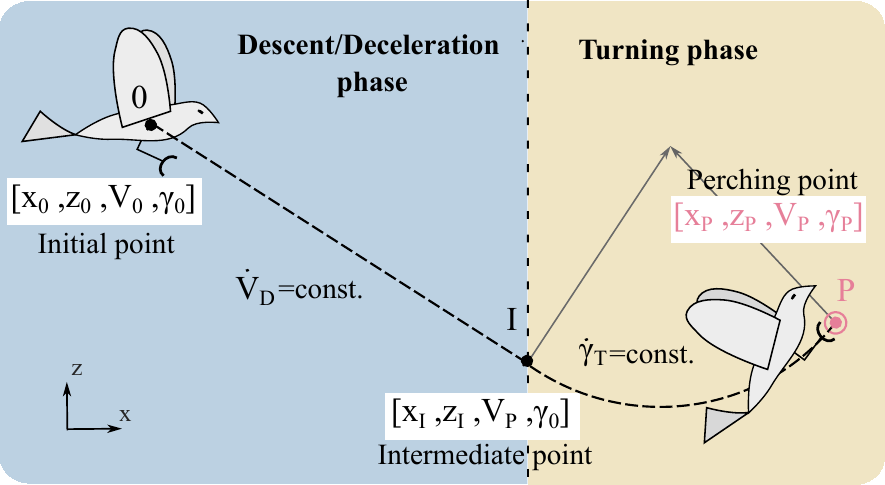}
    \caption{Trajectory model for approach maneuver. Three principal points (Initial, Intermediate and Perching) and two phases (Deceleration and Turning). 
    }
    \label{fig:trayectory}
\end{figure}

\noindent{\bf Maneuver kinematics.}
Ornithological evidence indicates that there are several phases in the approaching  flight of birds \cite{mechanics_perching,hawk_perching}, generally a descent followed by a rapid pitch up, as discussed in section \ref{sec:intro}. The approaching trajectory proposed for perching is described with the aid of the Fig. \ref{fig:trayectory}. 
The trajectory starts from a initial position, velocity and attitude $({x}_0,{z}_0,V_0,\gamma_0)$, and ends in a perching location and attitude $(x_P,z_P,V_P,\gamma_P)$. Thus, a two-phases trajectory is proposed, shown in Fig. \ref{fig:trayectory}. In the first one, called the \emph{``Descent phase''}, the velocity of the ornithopter is reduced by a constant deceleration $\dot{V}_D$ while maintaining constant the path angle $\dot{\gamma}=0$. In the second one, called \emph{``Turning phase''}, an uniform vertical turn, with constant angular rate $ \dot{\gamma}_T$ and velocity $\dot{V}=0$, is used to accommodate the attitude at perch. 
Under these conditions, the kinematics provide a family of trajectories composed by a linear segment tangent to a circumference arc, and parametrized with $(V_P,\gamma_P,\dot{V}_D,\dot{\gamma}_T)$.
To simplify the approach, clearance is assumed in order to not take into accountthe floor.
By imposing kinematic continuity in the intermediate point, the maneuver holds just two degrees of freedom. Thus, integrating the kinematics \eqref{eom1}-\eqref{eom2} from $(x_0=0,z_0=0)$ to $(x_P,z_P)$ yield
\begin{align}
x_P&=\frac{\cos{\gamma_0}}{2 \dot{V}_D}(V_P^2-V_0^2)+ \frac{\sin{\gamma_P}-\sin{\gamma_0}}{\dot{\gamma}_T} V_P, \label{VD0} \\
z_P&=\frac{\sin{\gamma_0}}{2 \dot{V}_D}(V_P^2-V_0^2)-
\frac{\cos{\gamma_P}-\cos{\gamma_0}}{\dot{\gamma}_T} V_P, \label{GT0}
\end{align}
with the total time
$
T_P=({V_P-V_0})/{\dot{V}_D}+({\gamma_P-\gamma_0})/{\dot{\gamma}_T}.
$
Denote the mean attitude as $\gamma_M:=(\gamma_P+\gamma_0)/2$, the vertical turn radius as $R_T$ and the mean characteristic distance $L_M$. It is straightforward to see that by rearranging the equations one can isolate the key accelerations which will be used later
\begin{equation}
\dot{V}_D=\frac{V_P^2-V_0^2}{2 L_{M}(\gamma_{P})}, \label{VD} 
\end{equation}

\begin{equation}
\dot{\gamma}_T=\frac{V_P}{R_T(\gamma_{P})}, \label{GT}
\end{equation}
where $\tan{\gamma_M}=-\frac{\cos{\gamma_P}-\cos{\gamma_0}}{\sin{\gamma_P}-\sin{\gamma_0}}$ and with
\begin{align*}
L_{M}(\gamma_{P})&:=\frac{x_P \tan{\gamma_M}-z_P}{\cos{\gamma_0} (\tan{\gamma_M}-\tan{\gamma_0})}, \\
R_T(\gamma_{P})&:=\frac{z_P-x_P \tan{\gamma_0}}{(\sin{\gamma_P}-\sin{\gamma_0})(\tan{\gamma_M}-\tan{\gamma_0})}.
\end{align*}
 It is worth noting that just $\{x,z,\dot{x},\dot{z},V,\gamma\}$ have assured continuity at the intermediate point. The discontinuity in the first derivative of $V_D$ and $\gamma_T$ by design can produce instabilities in the controller passing through this point, however the discontinuity is assumed small  enough to be physically possible, which is verified in simulation. Importantly, the adaptive controller designed can deal with these discontinuities.

\smallskip
\noindent{\bf Perching feasibility.}
\begin{figure}[tbp]
    \centering
    \includegraphics[width=0.95\columnwidth]{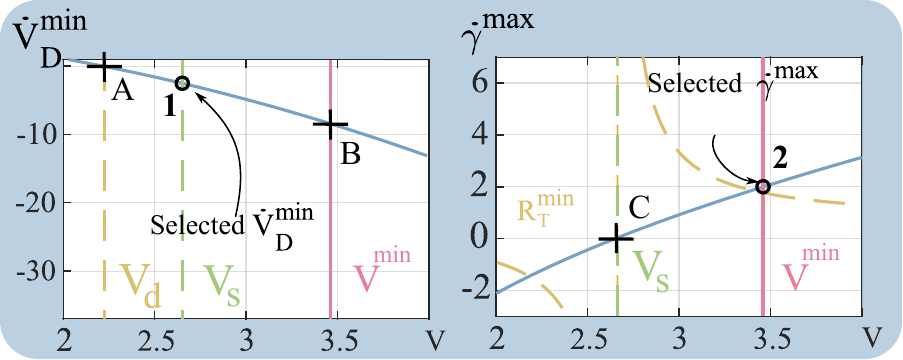}
    \caption{Selection of the minimum deceleration rate (left) and maximum turn rate (right). $V_d$ is the minimum velocity for deceleration, where $\dot{V}_D=0$ at point A. Point B is the magnitude at minimum velocity. 
    }
    \label{fig:dynamic_limitation}
\end{figure}
Flight limitations are critical, more in FW-UAV due to its instability and flapping, providing a wide range of unachievable maneuvers that will cause inevitably loss of control. Perching adds three main limiting factors: 1) the low maneuverability at low velocities of these platforms compared to birds and even more to rotary-blade UAVs; 2) the continuity of the kinematical trajectory itself; and 3) the final required pose so that the claw mechanism function properly at the perching location.

By physical motivation, and withouth any loos of generality, let assume $\gamma_0<0$ and hence $\gamma_P>\gamma_0$, since the perching point is below the cruise height and the pose must be positive, we decelerate to perch $V_P<V_0$ and $(x_0,z_0)=(0,0)$ for convenience.
The feasibility conditions depend on the aerodynamic model, mass properties, and mechanisms (see Appendix \ref{ap:aerodynamics}), and are enumerated below
\begin{itemize}
    \item [F1.]\textbf{Minimum velocity} ${V_P^{min}}$ is defined by the stall velocity multiplied by a safety factor. In particular, in ours $V_P^{min}= 1.3V_{s}\approx 3.5 \hspace{2pt} \frac{m}{s}$, with the stall velocity $V_{s}:=\sqrt{\frac{2mg}{\rho S C_{Lmax}}}$, for $C_{Lmax}=3.45$ at $\alpha_{s}=36 \hspace{2 pt}deg$.
    
     \item [F2.]\textbf{Minimum deceleration rate} ${\dot{V}_D^{min}}$. 
     Assuming the critical case in \eqref{eom3}, that is $\gamma:=\gamma^{min}=-37 \hspace{2 pt} deg$, we obtain a relation between the velocity and the minimum deceleration (for $C_D=C_D^{max}=2.95$, $\alpha=\alpha_{s}$ and $k=1$) shown in Fig. \ref{fig:dynamic_limitation}. We choose the minimum deceleration rate at stall velocity (point 1 in Fig.~\ref{fig:dynamic_limitation}), with a safety factor, that is $\dot{V}_D^{min}\approx -2 \hspace{2pt} \frac{m}{s^2}$. 
     
     \item [F3.] \textbf{Maximum turn rate} ${\dot{\gamma}_T}$, that is related to the maximum lift coefficient. For reference we choose the critical case from \eqref{eom4} at $\gamma=0$. At the bottom Fig. \ref{fig:dynamic_limitation} is presented the maximum turn rate as a function of the velocity. We choose this limit at $V=V_P^{min}$ (point 2 in Fig. \ref{fig:dynamic_limitation}), i.e. $\dot{\gamma}_T^{max}\approx 2\frac{rad}{s}$.
     
     \item [F4.] \textbf{Path geometric continuity} ${\gamma_P^{kin,min/max}}$. Enforcing the path to be continuous and tangent at the intermediate point $I$ (see Fig. \ref{fig:trayectory}), can be translated into a limitation in the reachable perching path angle.
     It can be demonstrated that given $\gamma_0$ a feasible solution of the maneuver only exists if $\gamma_P$ belongs to the range
\begin{align*}
        &[\gamma_P^{kin,min},\gamma_P^{kin,max}] 
        =[2 {\rm atan}{\frac{z_P}{x_P}}-\gamma_0,{\rm atan}{\frac{z_P}{x_P}}+\pi].
\end{align*}

    \item [F5.] \textbf{Mechanical perching pose} ${\gamma_P^{mec,min/max}}$.
The claw's performance limits the pose at perch. In particular, in ours the range is 
$\gamma_P  \in [10,60]deg.$
\end{itemize}
Note that F2-3 are simplified magnitudes of the maneuverability of the ornithopter, and they facilitate extrapolation to other ornithopters. However they are "rough magnitudes" 
, and they should be verified in simulation or experiments. 
For practical considerations, in what follows we consider only the set of feasible solutions, namely $\mathcal{Y}$, satisfying F1--F5.

\smallskip
\noindent{\bf Optimal solution for minimal velocity.}
The literature suggests that--during perching maneuvers--birds do not optimize either flight time or power consumption, since trajectories would be simply linear instead of concave as experiments show \cite{hawk_perching}. Moreover, authors in \cite{hawk_perching} demonstrate that the objective function depends on the minimal stall distance, which is extremely complicated to measure flying outdoors, where MCS are not available (recall {\bf R0}). However, this parameter is related to minimal velocity, which is crucial to minimize impact force and avoid prototype damage. This can be accurately measured in flight, and therefore, it is an excellent candidate to be used as the objective/cost function, as it is simpler and more feasible in practice than the minimum stall distance. 
Hence given the initial parameters ${\bf y_0}:=(\gamma_0, V_0, x_P, z_P)$, the optimization problem for the terminal velocity $V_{P}$, with the unknown parameters defined as ${\bf y}:=(\gamma_P, V_P,\dot{V}_D, \dot{\gamma}_T)$, is stated as follows
\begin{mini!} |s|
{\bf y\in \mathcal{Y}}{V_P}
{\label{J}}{}
\addConstraint{g_{1}:=}{\ \textnormal{Equation} \quad}{(\ref{VD0})\label{g1}}
\addConstraint{g_{2}:=}{\ \textnormal{Equation} \quad}{(\ref{GT0})\label{g2}}
\addConstraint{g_{3}:=}{\ \dot{V}_D^{min}-\dot{V}_D}{\leq0\label{g3}}
\addConstraint{g_{4}:=}{\ \dot{\gamma}_T-\dot{\gamma}_T^{max}}{\leq0\label{g4}}
\addConstraint{g_{5}:=}{\ \gamma_P^{min}-\gamma_P}{\leq0\label{g5}}
\addConstraint{g_{6}:=}{\ \gamma_P-\gamma_P^{max}}{\leq0\label{g6}}
\addConstraint{g_{7}:=}{\ V_P^{min}-V_P}{\leq0\label{g7}}.
\end{mini!}
Thus, it is a constrained optimization problem with 4 unknowns, 2 equality and 5 inequality constraints, namely $g_i$, $i=1,...,7$. Note that $g_{5,6}$ are fused from F4 and F5 such that $\gamma_P^{a}\in(\gamma^{b,a},\gamma^{b,a})$ with $a=min/max$ and $b=kin/mec$. Transforming the inequalities \eqref{g3}--\eqref{g7} to equality constraints by means of the slack variables $\sigma_i \in \mathbb{R}$, the constrained optimization problem \eqref{J} is solved by the Lagrange's multipliers method.
Hence, Lagrangian becomes
\begin{align*}
L({\bf y},\lambda,\sigma):=V_P\quad+\sum_{\substack{i=1 \\ j=i-2, \ i\geq 3}}^{7}\lambda_i \ g_i({\bf y},\sigma_j),
\end{align*}
where $\lambda\in\mathbb{R}^{7}$ and $\sigma\in\mathbb{R}^{5}$. The sufficient condition for optimality yields a search of solutions in a system with 16 equations and 16 unknowns given by
\begin{equation*}
\frac{\partial L({\bf y},\lambda,\sigma)}{\partial\bf y} = {\bf 0}_4, \quad \frac{\partial L({\bf y},\lambda,\sigma)}{\partial \sigma}= {\bf 0}_5, \quad g_i({\bf y},\sigma_j) = 0.
\end{equation*}
 The corresponding detailed equations are in the Appendix \ref{ap:solutions}. By analyzing the first set of 9 equations, it allows 6 possible solutions 
which have been calculated analytically.

The rest of the subsection is devoted to the analytical solution for all cases. 

\begin{itemize}
\item \textbf{Case 1:} $V_P=V_P^{min}$, ${\lambda_{1-6}=0}$\footnote{Note that the $\lambda$ not pointed out, e.g. $\lambda_7$ in Case 1, are  different from zero.}, ${\sigma_5=0}$.
This is of special relevance because it corresponds to the absolute minimum. When this is not the solution, it is necessary to solve the others for relative minima located at the boundaries, and in case of several relative optima we select the one with the lowest $V_{P}$ possible. 
From \eqref{VD} and \eqref{GT} the constraints $g_{1-4}$ read
\begin{align}
\dot{V}_D^{min} - \frac{(V_P^{min})^2-V_0^2}{2L_{M}(\gamma_{P})} &\leq 0,  \label{eqC11} \\
\frac{V_P^{min}}{R_{T}(\gamma_{P})} - \dot{\gamma}_T^{max} &\leq 0, \label{eqC12}
\end{align}
which is a system of two inequalities and one variable $\gamma_P$. Thus, this case has a solution if there exist and overlap between the range of solutions of \eqref{eqC11}--\eqref{eqC12} and the range of feasible attitudes for perching of \eqref{g5}--\eqref{g6}. The equation (\ref{eqC11}) is transformed to
\begin{equation*}
   \frac{\tan{\gamma_M}-\tan{\gamma_0}}{\tan{\gamma_M}-z_P/x_P } \leq \frac{2 x_P \dot{V}_D^{min}}{((V_P^{min})^2-V_0^2) \cos{\gamma_0}}=:\Xi.
\end{equation*}
It is straightforward to see, that if the right hand side of the inequality $\Xi>1$, the solution reads 
\begin{equation*}
  \gamma_P^{*} > 2 \arctan{\left(  \frac{\tan{\gamma_0}-\Xi \ z_P/x_P }{1-\Xi}  \right)}-\gamma_0=:\gamma_P^{c1,min}
\end{equation*}
However, if $\Xi \in (0,1)$ the inequality are inverted.
On the other hand, the equation (\ref{eqC12}) corresponds to
\begin{align*}
  \gamma_P^{*} <& \ 2 \arcsin{\left( \frac{\dot{\gamma}_T^{max}}{2 V_P^{min}}  \cos{\gamma_0} (z_P-x_P \tan{\gamma_0}) \right)} \\
  &+\gamma_0=:\gamma_P^{c1,max}
\end{align*}
Therefore, the solution is defined through the range 
$$
\gamma_P^*\in (\gamma_P^{c1,min},\gamma_P^{c1,max})  \cap  (\gamma_P^{min},\gamma_P^{max}).
$$ 
In general the solution is an interval and so, additional constraints could be proposed to reduce $\gamma_P^*$ to a single point. However, for simplicity we select the midpoint of such interval. Once $\gamma_P^{*}$ is calculated, \eqref{eqC11} and \eqref{eqC12} provides the optimal solution of $\dot{V}_D^{*}$ and $\dot{\gamma}_T^{*}$.

\item \textbf{Case 2:} ${{\dot{\gamma}_T=\dot{\gamma}_T^{max}}}$, ${{\gamma_P=\gamma_P^{max}}}$, ${\lambda_{3,5,7}=0}$ , ${\sigma_{2,4}=0}$.

The solution is obtained directly by isolating the two remaining unknowns from \eqref{g1}--\eqref{g2} that yield \eqref{VD}--\eqref{GT} and imposing $\dot{\gamma}_T^{max}$ and $\gamma_P^{max}$. Thus, from \eqref{GT} and \eqref{VD}, respectively, we get

\begin{equation*}
 V_P^*=\dot{\gamma}_T^{max} R_{T}(\gamma_P^{max}) \Longrightarrow
 \dot{V}_D^*=\frac{(V_P^*)^2-V_0^2}{2 L_{M}(\gamma_P^{max})}.
\end{equation*}
 By assumption in this case, the equations \eqref{g3} and \eqref{g7} are free and hence, the above solution is feasible with $V_P^*>V_P^{min}$ from \eqref{g7} and $\dot{V}_D^*>\dot{V}_D^{min}$ from \eqref{g3}. 

\begin{figure*}[t]
    \centering
    \includegraphics[width=\textwidth]{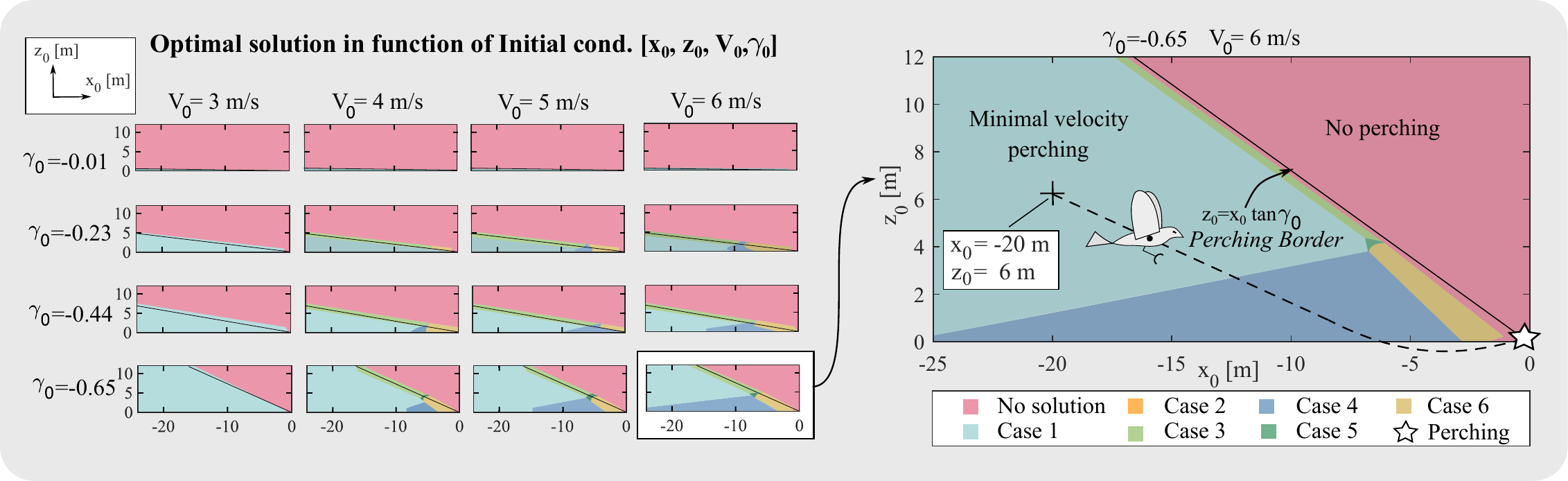}
    \caption{Optimal case for the minimum velocity trajectory as a function of initial position
, attitude and velocity. Perching point $\star$ at $(x_P,z_P)=(0,0)$ at the bottom-right. Left: the results at constant initial attitude $\gamma_0$ and velocity $V_{0}$ for several values. Right: the optimal solution for $\gamma_0=-0.65 \ rad$ and $V_0=6 \ m/s$ is zoomed in, and an sample trajectory defined by the initial point is highlighted.}\label{fig:perching_optimal}
\end{figure*}

\item \textbf{Case 3:} $\dot{\gamma}_T=\dot{\gamma}_T^{max}$, $\gamma_P=\gamma_P^{min}$, ${\lambda_{3,6,7}=0}$, ${\sigma_{2,3}=0}$.\\
It is solved in a similar way to {Case 2}. 

\item \textbf{Case 4:} ${\dot{V}_D=\dot{V}_D^{min}}$,${\gamma_P=\gamma_P^{max}}$, ${\lambda_{4,5,7}=0}$, ${\sigma_{1,4}=0}$. 

The solution is obtained directly by isolating the two remaining unknowns from \eqref{g1}--\eqref{g2} that yield \eqref{VD}--\eqref{GT} and imposing $\dot{V}_D^{min}$ and $\gamma_P^{max}$. Thus, from \eqref{VD} and \eqref{GT}, respectively, we get
\begin{equation*}
 V_P^*=\sqrt{2 \dot{V}_D^{min} L_{M}(\gamma_P^{max}) + V_0^2} \Rightarrow \dot{\gamma}_T^* = \frac{V_P^*}{R_{T}(\gamma_P^{max})}. 
\end{equation*}
 The equations \eqref{g5} and \eqref{g7} are free and hence, the solution is feasible if it satisfies $V_P^*>V_P^{min}$ from \eqref{g7} and $\dot{\gamma}_T^*<\dot{\gamma}_T^{max}$ from \eqref{g4}.  

\item \textbf{Case 5:} ${\dot{V}_D=\dot{V}_D^{min}}$,${\gamma_P=\gamma_P^{min}}$, ${\lambda_{4,6,7}=0}$, ${\sigma_{1,3}=0}$.
It is solved in a similar way to {Case 4}. 

\item \textbf{Case 6:} ${\dot{V}_D=\dot{V}_D^{min}}$,${\dot{\gamma}_T=\dot{\gamma}_T^{max}}$, ${\lambda_{5,6,7}=0}$, ${\sigma_{1,2}=0}$. 
This case is more involved, because it is necessary to isolate $V_P$ and $\gamma_P$ from \eqref{g1} and \eqref{g2} with \eqref{VD} and \eqref{GT}. 
To this end, let us define the following non-dimensional positive-definite parameters:
\begin{equation*}
\bar{\dot{V}}_D=-\frac{2\dot{V}_D}{x_P  \dot{\gamma}_T^2}, \  
\bar{V}_0= \frac{V_0}{x_P  \dot{\gamma}_T}, \ 
\bar{z}=z_P/x_P-\tan{\gamma_0}.
\end{equation*}
Defining the change of variables given by ${\bf a}:=\tan{\gamma_M}-\tan{\gamma_0}$, and ${\bf b}:= \frac{V_P}{x_P \dot{\gamma}_T} $ and plugging the parameters above, the equations to be solved 
becomes a fourth-degree polynomial in ${\bf a}$  with no analytical solution found, where 
all the coefficients except the first one have a definite sign $(? \ + \ - \ + \ -)$. Thus, if the first coefficient is positive Descartes' rule of signs ensures at least one real root.
 The equations \eqref{g5}, \eqref{g6} and \eqref{g7} are free and hence, the solution is feasible if it satisfies $V_P^*>V_P^{min}$ from \eqref{g7} and  $\gamma_P^* \in (\gamma_P^{min},\gamma_P^{max})$ from \eqref{g5}--\eqref{g6}.  
\end{itemize}

\noindent{\bf Optimal solution analysis.} 
In Fig. \ref{fig:perching_optimal} the results of the optimal trajectories are shown, for a wide range of initial conditions $(x_0,z_0,V_0,\gamma_0)$. Recall that the reference position has been translated to $(x_P,z_P)=(0,0)$. If the ornithopter initial point is located in the red zone (no solution), perching is not possible. This zone is mainly located at $\tan{\gamma_0}<\frac{z_0}{x_0}$, which defines a zone behind a line with $\gamma_0$ slope through the perching point, henceforth, the \textit{Perching Border}. At an altitude under this line, the perching is possible at the minimum velocity (\textbf{Case 1}) except in some cases near the perching point at maximum and minimum altitudes. In such cases, the minimum velocity reached in perching is higher than the ornithopter minimal velocity, i.e. sub-optimal, this is any initial point outside the light blue zone (\textbf{Case 1}). 

Therefore, it should be noted that the more negative the initial trajectory angle, the easier to reach the minimum perching velocity. Of course, the minimal velocity is reached easily as initial velocity is smaller. Near the \textit{Perching Border} and perching point,  just \textbf{Case 6} is possible, that is an agressive maneuver where the accelerations are at the limit. This zone should be avoided to perch since it demands the ornithopter to be near the dynamic limits. 
\begin{figure}[htbp]
    \centering
\includegraphics[width=0.85\columnwidth]{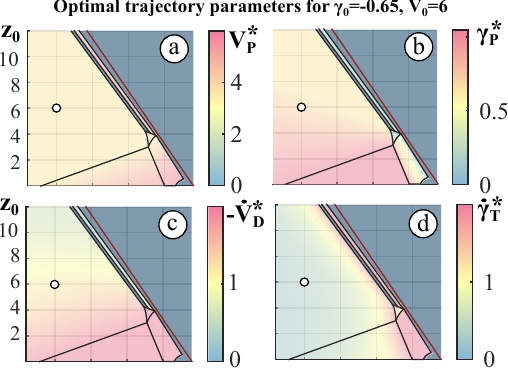}
    \caption{Optimal trajectories for $\gamma_0=-0.65 \ rad.$ and $V_0=6 \ m/s$. The sample point  highlighted as $\circ$ corresponds to: $x_0=-20$ $m$, $z_0=6$ $m$, $\dot{V}_D^*=-1.15$ $m/s^2$, $\dot{\gamma}_T^*=0.37$ $rad/s$, ${V}_P^*=3.5$ $m/s$ and $\gamma_P^*=0.69$.} 
    \label{fig:optimal}
\end{figure}
For the same cases, the optimal trajectory parameters are shown in Fig. \ref{fig:optimal}. In Sub-figure (a) the perching velocity is presented, which is increasing from the minimum reachable as the perching point is closer, that is from yellow to red gradient in the figure Similar performance exhibits the optimal perching angle in (b), deceleration rate (c), increasing as perching point is closer. The turning rate (d) increases as the initial point is closer to the \textit{Perching Border} mentioned before. This is due to the space restriction, which forces the trajectory to be more aggressive.\\ 
The validation with real hawk's flight of the proposed optimal perching solutions are presented in Section \ref{sec:Results}.

\section{Nonlinear Control}\label{controller}
In contrast to classical UAV, the ornithopter dynamics \eqref{eom3}--\eqref{eom6} 
are nonlinear and non autonomous, since it includes the time dependence of flapping and control signals. Moreover, it is 
under-actuated with 2 inputs $\{k, \delta_{e}\}$ and 4 output states $\{V, \theta, \gamma, q\}$. Inaccuracies are expected due to its complexity, notice the difference with a quad-copter in \cite{carlos_aero} where we could completely characterize the steady dynamics in standard operational regimes.
Additionally, from our know-how of adaptive nonlinear control in different fields \cite{gavilan2015}, we propose it to control the path, as discussed in Section \ref{sec:intro}. 
However, the aerodynamic model derived in Appendix \ref{ap:aerodynamics} is still quite complex for the control design, so we reduce its order while maintaining accuracy at least 98\% RMSE just for this purpose (note that for simulation the full model is used). 
The dependencies of the reduced model are: 
\begin{align*}
\Theta_{{D}_w}^T r_w & \approx
(\Theta_{{D}_w}^r)^{T} r_{{D}_w} :=
(\Theta_{{D}_w}^r)^{T} \left[1, \alpha^2, k^4 \right]^T, \\
\Theta_{{L}_w}^{T} r_w & \approx 
(\Theta_{{L}_w}^r)^{T} r_{{L}_w} :=
(\Theta_{{L}_w}^r)^{T} \left[1, \alpha, \alpha^2, k^2, \alpha k \right]^T.
\end{align*}
The upperscript $(\cdot)^r$ stands for `reduced', due to the lower dimension of the aerodynamic coefficients vector. However, for simplicity we will omit it in the rest of the section.

On the other hand, the velocity equation \eqref{eom3} is simplified taking into account the magnitude orders of $F_X$. Thus, while $|{D_w}| \sim 10^1$, for the other aerodynamic factors $|{L_t \sin (\alpha_t-\alpha)}| \sim |{D_t \cos (\alpha_t-\alpha)}|\sim 10^{-1}$ and $D_b \sim 10^{-2}$, so they have been neglected for the control design model.

\subsection{Velocity control: $\Sigma_{V}$}
For the sake of clarity, let define $\bar{q}=\frac{\rho S}{2m}$ and split the definition of the wing's drag coefficient above as $\Theta_{{D}_w}^{T} r_{{D}_w} = (\Theta_{{D}_w}^{\alpha})^{T} r_{{D}_w}^\alpha+\Theta_{{D}_w}^{k} r_{{D}_w}^k$ where $r_{{D}_w}^\alpha=(1, \alpha^2)^T$ and $r_{{D}_w}^k=k^4$, that is, the contribution of drag and thrust, respectively. 
Consider the velocity dynamics \eqref{eom3} together with \eqref{F1}  
and define the regulation error as $e:=V-V_R$, where $V_R$ is the reference velocity. The error dynamics yield
\begin{equation}
\dot{e}=-\Bar{q}(e+V_R)^2(\Theta_{{D}_w}^\alpha)^{T} r_{{D}_w}^\alpha+C_T-g\sin{\gamma}-\dot{V}_R, \label{eV}
\end{equation}
where the control action has been compactly defined as the thrust generation by the wing given by 
\begin{equation}
C_T:=-\Bar{q}(e+V_R)^2\Theta_{{D}_w}^k r_{{D}_w}^k.\label{ct_k}
\end{equation}
The fact that \eqref{eV} is scalar eases the controller design via Lyapunov. Thus, the adaptive feedback proposed is
\begin{align}
C_T&=- k_0 e + \Bar{q}V_R^2(\hat{\Theta}_{{D}_w}^\alpha)^{T} r_{{D}_w}^\alpha+g\sin{\gamma}+ \dot{V}_R , \label{control_ct} \\
\dot{\hat{\Theta}}_{{D}_w}^\alpha &= - \Bar{q} V_R^2 e   (r_{{D}_w}^\alpha)^{T} \Gamma_V  \label{control_ct_ad}
\end{align}
where $\hat{\Theta}_{{D}_w}^\alpha$ is the estimate of $\Theta_{{D}_w}^\alpha$ for the adaptive scheme with the error defined as $\tilde{\Theta}_{{D}_w}^\alpha:={\Theta}_{{D}_w}^\alpha-\hat{\Theta}_{{D}_w}^\alpha$, and $k_0$ is a positive control gain.

The stability guarantees are derived with the following positive definite Lyapunov function
\begin{equation}
W:=\frac{1}{2}e^2 + \frac{1}{2} (\tilde{\Theta}_{{D}_w}^\alpha)^{T} \Gamma_V^{-1} \tilde{\Theta}_{{D}_w}^\alpha, \label{W}
\end{equation}
where $\Gamma_V$ is the adaptation gain defined as a positive define $(2\times 2)$-matrix. Thus, the derivative of $W$ along the trajectories of \eqref{eV} reads 
\begin{align*}
\dot{W}&=e\dot{e} - (\tilde{\Theta}_{{D}_w}^\alpha)^{T} \Gamma_V^{-1} \dot{\hat{\Theta}}_{{D}_w}^\alpha \\
&= -\bar{q}e^2(e+2V_R)(\Theta_{{D}_w}^\alpha)^{T} r_{{D}_w}^\alpha -k_0 e^2 \leq -k_0 e^2,
\end{align*}
where we have included the proposed controller \eqref{control_ct} and \eqref{control_ct_ad} 
unfolded the term $(e+V_R)^{2}$ and used $\dot{\tilde{\Theta}}_{{D}_w}^\alpha=-\dot{\hat{\Theta}}_{{D}_w}^\alpha$. 
The last inequality is obtained noting that $(\Theta_{{D}_w}^\alpha)^{T} r_{{D}_w}^\alpha>0$ by construction, and so does $e+2V_R>0$ because $V>-V_R$ is always true.
We state the stability result in the proposition below.
\begin{prop} \label{pr:vel}
Consider the velocity error dynamics \eqref{eV} and $V_{R}(t), \dot V_{R}(t), \ddot V_{R}(t)\in \cal{L}_\infty$. For any control gain $k_0$ and $\Gamma_V$ positive, the adaptive state feedback \eqref{control_ct}--\eqref{control_ct_ad} guarantees the boundedness of $e$ and ${\hat{\Theta}}_{{D}_w}^\alpha$ and the convergence of $e$ to zero, i.e. $\lim_{t\to \infty} |e(t)| = 0$.
\end{prop}
\begin{proof}
The function \eqref{W} is bounded by $\underline{\nu}(e,\tilde{\Theta}_{{D}_w}^\alpha) \leq W \leq \overline{\nu}(e,\tilde{\Theta}_{{D}_w}^\alpha)$ for some $\underline{\nu},\overline{\nu} \in \mathcal{K}_{\infty}$, and hence it is positive definite radially unbounded. From the derivations above $\dot W\leq 0$ and $V_{R}, \dot V_{R}\in \cal{L}_\infty$ and therefore, we conclude global boundedness of $e$ and the estimate ${\hat{\Theta}}_{{D}_w}^\alpha$.
To prove convergence first note that the closed loop \eqref{eV} is non autonomous. On the one hand, $W$ is bounded from below and $\dot W \leq 0$, hence $e \in {\cal{L}}_2 $. On the other hand, the boundedness of $e$ and $\ddot V_{R}$ imply that $\ddot W \in \cal{L}_{\infty}$ and therefore $e(t)$ is uniformly continuous. Finally, Barbalat's lemma guarantees that $\lim_{t\to \infty} |e(t)| = 0$.
\end{proof}
Finally, by extracting the variable implementation from \ref{control_ct}, the control action leads to
\begin{align*}
 f^4=\Big(  -{\Dot{V}_R }  
-{\Bar{q} V_R^2  \hat{\Theta}_{{D}_w}^\alpha r_{{D}_w}^\alpha }
-{g \sin{(\gamma)}}  + k_0 e \Big) \frac{V^2}{{\Bar{q} \Theta_{{D}_w}^{k} (\pi c)^4 }} \nonumber 
\end{align*}
where the adaptive dynamics were defined in \ref{control_ct_ad} with an initial condition ${{\hat{\Theta}}_{{D}_w}^{\alpha}}|_{t=0}={{{\Theta}}_{{D}_w}^{\alpha}}$
And finally the parameters to tune: $k_0>0$, $\Gamma_V>0$.
\subsection{Attitude control: $\Sigma_{\gamma}$}

The attitude dynamics are described by the equations \eqref{eom4}--\eqref{eom6}, and the control is developed by using the adaptive backstepping methodology. Let us first set some necessary theoretical background. 

\smallskip
\noindent{\bf General Averaging.} Certainly, the attitude dynamics is strongly dominated by the flapping aerodynamics of \eqref{F2} which are non autonomous mainly due to the
flapping (see Appendix \ref{ap:aerodynamics}) that cannot be disregarded and neglected unlike 
in fixed-wing aircraft. 
However, for our prototype, the phugoid and short-period modes have natural frequencies about an order of magnitude lower than the lowest flapping frequency during the maneuver \cite{Lanchester,TAYLORTHOMAS2002}.
Therefore, such model satisfies the necessary requirements for the general averaging method, which includes also non-periodic well-behaved bounded average functions (see e.g. the seminal book \cite{khalil}). In essence, averaging relates the solution for the averaged dynamical system with that of the original periodic-like system. Let change coordinates and time scale to better see that the dynamics \eqref{eom4}--\eqref{eom6} satisfies the averaging method requirements. Thus, define a new time scale $\tau:=\omega t$, $\omega>1$, the dimensionless parameter $\epsilon:=I_y/(m c^2 \omega)$---for our ornithopter $\epsilon \approx 0.53/\omega$---and the new coordinates $\chi_1:=\omega^2 I_y \epsilon^{2} \theta$, $\chi_2:= \omega^2 I_y \epsilon \mathring{\theta}$ and $\chi_3:=\omega m \epsilon \gamma$, where $\mathring{(\cdot)}$ stands for the time derivative in the new scale. Hence, let $\chi:=(\chi_1, \chi_2, \chi_3)^{T}$, the attitude dynamics in these new coordinates and time scale become
\begin{equation} \label{avg}
\mathring{\chi}= \epsilon \left(\begin{array}{c}\chi_{2} \\F_{M} \\ (-m g + F_{Z})/V \end{array}\right) =: \epsilon h(\chi, \tau, \epsilon).
\end{equation}
Notice that the dynamics are in the standard averaging form with the small parameter $\epsilon$ perturbing the evolution of $h(\chi, \tau, \epsilon)$, which means that for $0<\epsilon<1$ the changes in $\chi$ are much slower than the changes due to the flapping. We summarize the stability result in the following proposition, conveniently adapted from \cite{khalil}. However, for the sake of simplicity, avoiding heavy notation, we omit some technicalities, retaining only those strictly necessary, and refer to interested readers to \cite{khalil} for all the technical details.
\begin{prop}
Consider dynamics \eqref{avg} and define the corresponding average dynamics as $\label{T-avg}
\mathring{\chi} = \epsilon h_{av}(\chi)$, with the average function defined as
\begin{equation*} 
h_{av}(\chi):=\lim_{T \to \infty} \frac{1}{T}\int_{t}^{t+T} h(\mu, \chi, 0) d \mu, \ T>0,
\end{equation*}
both \eqref{avg} and \eqref{T-avg} twice differentiable and bounded in every compact set of the $\chi$-domain $\mathcal{D} \subset \mathbb{R}^{3}$. 
Let $\chi(\tau,\epsilon)$ and $\chi_{av}(\epsilon\tau)$ denote the solutions of \eqref{avg} and \eqref{T-avg}, respectively. If $\chi_{av}(\epsilon\tau)\in \mathcal{D}$ for all $\tau\in[0,\zeta/\epsilon]$, $\zeta\geq 0$, and $\chi(0,\epsilon) - \chi_{av}(0)=\mathcal{O}(\nu(\epsilon))$, then there exists an $\epsilon^{*}>0$ such that for all $0<\epsilon<\epsilon^{*}$, $\chi(\tau,\epsilon)$ is well defined and
$$
\chi(\tau,\epsilon) - \chi_{av}(\epsilon\tau) = \mathcal{O}(\nu(\epsilon)) \ \textnormal{on} \ \tau \in [0, \zeta/\epsilon],
$$
for some function $\nu\in \mathcal{K}$.
\end{prop}

\begin{remark}
The result above formally ensures what the intuition tells us. In simple words, if we design the controller with the averaged attitude dynamics instead of the time-dependent ones, the convergence is guaranteed and  depends on $\epsilon$.
\end{remark}
In what follows, for the controller design we consider the average dynamics of \eqref{eom4}--\eqref{eom6}, that corresponds to \eqref{T-avg}, with the average of \eqref{F2}.

\begin{figure}[htbp]
    \centering
\includegraphics[width=0.95\columnwidth]{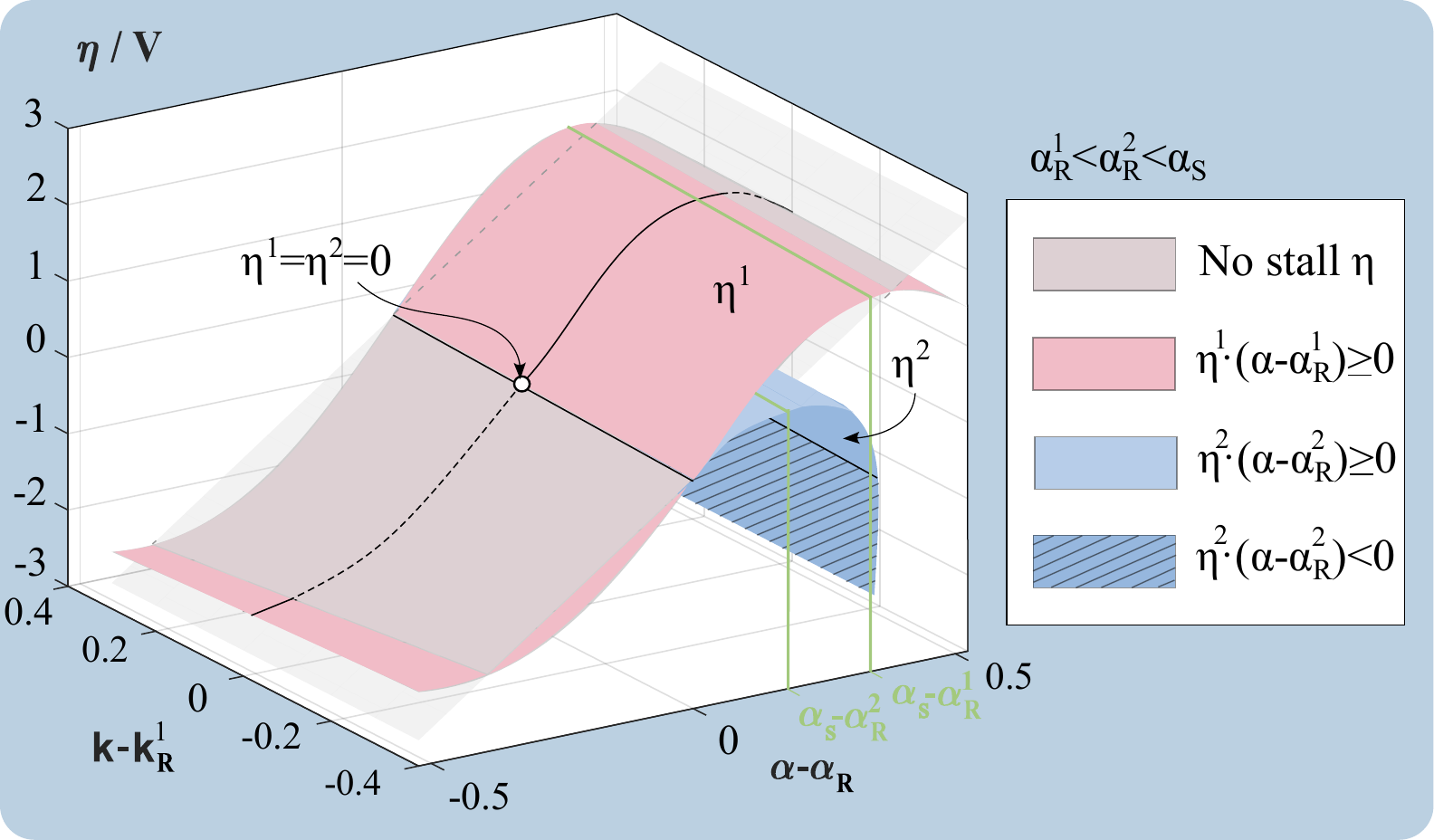}
    \caption{Function $\eta^1=\eta(\alpha_R^1)$ for $\alpha_R^1=10 \hspace{2pt}deg$ and $k_R=1.2$ and $\eta^2$ for $\alpha_R^2=20 \hspace{2pt}deg$ Note that $\eta (\alpha-\alpha_R) \geq 0$ for $\alpha_R<\alpha_s$ even if $\alpha$ is greater than the stall ($\alpha_s$) as $\alpha_R<\alpha_s$.}
    \label{fig:eta}
\end{figure}

\smallskip
\noindent{\bf Adaptive backstepping.} Similarly, as in the velocity control case, 
let the error variables be defined as
$ \label{e_gamma}
    e_1:=\gamma-\gamma_R, \quad
    e_2:=\theta-\theta_R, \quad
    e_3:=q-q_R,
$
 where subscript $(\cdot)_R$ stands for the reference states. Let assume the path is smooth,  $\dot{q}_R \approx 0$ and $\cos{\gamma} \approx \cos{\gamma_R}$. 
Thus, the error dynamics from \eqref{eom4}--\eqref{eom6} can be rewritten as
\begin{align}
	\dot{e}_1&=\eta(\alpha), \label{eqe1}\\
    \dot{e}_2&= e_3, \label{eqe2}\\
    \dot{e}_3&= \beta_{2w} \mu(\alpha) V^2 (\Theta_{{L}_w})^{T} r_{{L_w}}(\alpha) \cos{\alpha} x_a \nonumber \\ 
    &- \beta_{2t}V^2 \Theta_{{L}_t}^1 \sin{\big(\Theta_{{L}_t}^2 u\big)}, \label{eqe3}
\end{align}
where we have defined
\begin{equation} \label{eta}
\eta(\alpha) := \beta_1 V \mu(\alpha) (\Theta_{{L}_w})^{T} r_{{L_w}}(\alpha) -g \cos{\gamma_R}/V -\dot{\gamma}_R,
\end{equation}
and the parameters $\beta_1:=\frac{\rho S}{2m}$, $\beta_{2w}:=\frac{\rho S}{2I_{y}}$ and $\beta_{2t}:=\frac{\rho S_t}{2 I_{y}}$. Additionally, for compactness, the control action is defined as $u:=\delta_e + \alpha +\frac{x_t q}{V} $, that accounts the angle of attack induced in the tail due to the pitch motion. Next, we state a property of aerodynamic stability. To do this, let $\cal F$ denote the flight envelope of the ornithopter, which includes post-stall regime.
\begin{property} \label{P}
The trim point $\dot e_{1}=0$, from \eqref{eqe1} and \eqref{eta}, defines the angle of attack $\alpha_{R}$ such that $\eta(\alpha_{R})=0$. The following property holds: 
\begin{equation} \label{eq:P}
\forall \bar{e} \in {\cal F}, \ \exists \sigma>0: \ |\bar{e}|<\sigma \ \Longrightarrow \ \bar e \cdot \eta(\bar e + \alpha_{R})\geq 0.
\end{equation}
This is graphically described in Fig. \ref{fig:eta}, including the the stall.
\end{property}
\noindent In the following remark, we highlight the main difficulty of controller design in ornithopters compared to aircraft.

\begin{remark}
The control stability condition is much simpler in fixed-wing aircraft \cite{gavilan2015} since aerodynamics are linear throughout $\cal{F}$, and hence the property \eqref{eq:P} becomes true for $\sigma \to \partial \cal{F}$, with $\partial \cal{F}$ the boundary of $\cal{F}$. It is noteworthy that at perching manoeuvre in ornithopters, the flapping and stall regimes limit such range. 
Thus, this work extends the stability criterion to ornithopters perching, shown in Fig.~\ref{fig:eta}. This fact adds a hard requirement in the controller design.
\end{remark}

\noindent{\bf Step 1.} Let start with the subsystem (\ref{eqe1}) and define the virtual control $e_{2}=v_2:=-k_{1}e_{1}$. Consider the function $W_1:=e_1^2/2$ as a Lyapunov candidate. Its derivative yields
\begin{align*}
\dot{W}_1 &= 
  - \bar e \ \eta(\bar e +\alpha_{R})/(1+k_{1}),
\end{align*}
where $\bar e:=-(1+k_{1})e_1$. Property \ref{P} guarantees a range of the control gain $k_{1}$ for which $\dot W_{1}\leq 0$. Thus defining the parameters of wing aerodynamic component-wise as $\Theta_{{L}_w} := (\Theta^1, \Theta^\alpha, \Theta^{\alpha^2}, \Theta^{k^2}, \Theta^{\alpha k})^{T}$ the range yields
\begin{equation*}
    0 < 1+k_1 \leq \big({\Theta^{\alpha}+2\alpha_R^{min}\Theta^{\alpha^2}+k^{min}\Theta^{\alpha k}}\big) \big / \big({e_1^{max}\Theta^{\alpha^2}}\big). 
\end{equation*}
An estimate for our prototype with $e_1^{max}=60 \deg$, $\alpha_R^{min}=0.18$ and $k^{min}=0.64$ yields $1+k_1 \leq 0.84$. 
For ease of reference, we denote $k_1 \in \Omega_1$ with $\Omega_1 := (-1, -0.16]$.

\noindent{\bf Step 2.}
Define the manifold error as $\tilde{e}_2:=e_2-v_2=e_2+k_1 e_1$. Consider the subsystem formed by \eqref{eqe1} and the error dynamics $\dot{\tilde{e}}_2=e_3+k_1\eta$, from \eqref{eqe2}, with the virtual control $e_3=v_3$ to be defined further. 
Let define the Lyapunov function candidate 
\begin{align*}
W_2:=c_1 W_1 + \int_{\alpha_R}^{\Tilde{e}_2-(1+k_1)e_1+\alpha_R} \eta(s)  \,ds,
\end{align*}
which is positive-definite under the Property \ref{P} and with $c_1$ a positive constant. Its derivative becomes
\begin{align*}
\dot{W}_2 
&= (c_1 e_1 + v_3) \eta(\alpha) -\eta(\alpha)^{2},
\end{align*}
where with some abuse of notation, just for compactness, we have used $\alpha=\tilde e_{2}-(1+k_{1})e_1+\alpha_{R}$ from previous definitions. 
Therefore, defining the virtual control as $v_3:=-c_1 e_1$ enforces $\dot{W}_2 \leq 0$ and the stability of this subsystem. \\
\noindent{\bf Step 3.}
In this last step, we derive jointly the backstepping and adaptive control. To facilitate the controller design we make some arrangements in the dynamics. Thus, let define the final manifold error as $\tilde{e}_3=e_3-v_3=e_3+c_1 e_1$, and hence the error dynamics from (\ref{eqe3}) are given by
\begin{align}
\dot{\tilde{e}}_3&= V^2  \beta_{2w} \mu(\alpha) (\Theta_{{L}_w})^{T} r_{{L_w}}(\alpha) \cos{\alpha} x_a +c_1 \eta(\alpha) \nonumber \\ 
&- V^2  \beta_{2t} \Theta_{{L}_t}^1 \sin{\big(\Theta_{{L}_t}^2 u\big)}, \nonumber \\ 
&= V^{2} \beta_{2t} \Theta_{{L}_t}^1 \left[\Phi^{T} \Psi(\alpha, \tilde e_{3}) -\sin{\big(\Theta_{{L}_t}^2 u\big)} \right]\nonumber \\ 
& - V^{2} \beta_{2w} x_{a} k_3 \tilde e_{3} + c_1 \eta(\alpha).\label{e3_tilde_dot}
\end{align}
On the one hand, we have defined the measurable vector function $\Psi(\alpha, \tilde e_{3})=\Psi(\alpha, q, \gamma, \gamma_{R})$, that has been extended adding and subtracting $k_3 \tilde e_{3}$ to obtain an extra stabilizing negative term at the cost of an additional little complexity. On the other hand, we redefine the physical parameters, collecting them in the vector $\Phi$. Hence they are given by
\begin{align*}
\Psi(\alpha, \tilde e_{3})&:=\left(\begin{array}{c}\mu(\alpha) r_{{L_w}}(\alpha) \cos{\alpha} \\ k_3 \tilde e_{3}\end{array}\right) \in \mathbb{R}^{6}, \\
\Phi(\Theta_{L}) &:= \frac{\beta_{2w}x_{a}}{\beta_{2t} \Theta_{{L}_t}^1} \left(\begin{array}{c} \Theta_{{L}_w} \\ 1 \end{array}\right) \in \mathbb{R}^{6}.
\end{align*}
Let define the following Lyapunov function candidate as
\begin{equation} \label{W3}
W_3:=W_2 + \frac{1}{2 c_1}\Tilde{e}_3^2 + \frac{|\Theta_{{L}_t}^1|}{2} \tilde{\Phi}^{T} \Gamma_{\gamma}^{-1} \tilde{\Phi}, 
\end{equation}
where  $\tilde{\Phi}:=\Phi-\hat{\Phi}$ is the parameter error with $\hat{\Phi}$ the estimate and $\Gamma_{\gamma}>0$ is the adaptation matrix. Its derivative becomes
\begin{align}
\dot{W}_3 &= -\eta(\alpha)^{2} +2 \tilde{e}_3 \eta(\alpha) - V^{2} \beta_{2w} x_{a} \frac{k_3}{c_1} \tilde e_{3}^{2} \nonumber \\
&+ \frac{\tilde{e}_3}{c_1}V^{2} \beta_{2t} \Theta_{{L}_t}^1 \big[\Phi^{T} \Psi(\alpha,\tilde e_{3}) -\sin{\big(\Theta_{{L}_t}^2 u\big)} \big] \nonumber \\
&-|\Theta_{{L}_t}^1| \ \tilde{\Phi}^{T} \ \Gamma_{\gamma}^{-1} \ \dot{\hat{\Phi}}. \label{dW3}
\end{align}
Finally, we first define the static-state feedback as
\begin{equation} \label{u}
u:= \arcsin{(\hat{\Phi}^{T} \Psi(\alpha,\tilde e_{3})})/\Theta_{{L}_t}^2. 
\end{equation}
Thus, upon plugging \eqref{u} in \eqref{dW3} 
and removing the uncertaing by defining the adaptation law as
\begin{equation} \label{dPhi}
\dot{\hat\Phi}:= \Gamma_{\gamma} \frac{\beta_{2t}}{c_1} {\rm sign}({\Theta_{{L}_t}^1}) V^{2} \tilde{e}_3 \Psi(\alpha,\tilde e_{3}),
\end{equation}
where the sign function is defined as ${\rm sign}(\zeta):=1$ if $\zeta \geq 0$ and $-1$ elsewhere. The fact that $\dot W_3\leq 0$ can be seen with Young's inequality. Certainly, redefining the control gain as $\bar k_3:=V^{2} \beta_{2w} x_{a} k_3/c_1$, an upper bound for $\dot W_3$ becomes
\begin{align}
\dot{W}_3 &= -\eta(\alpha)^{2} +2 \tilde{e}_3 \eta(\alpha) - \bar k_3 \tilde e_{3}^{2} \nonumber \\
&\leq - (1-1/\lambda) \eta^2 -  (\bar{k}_3 -\lambda) \tilde{e}_3^2, \label{dW3b}
\end{align}
which is negative semidefinite for any $\bar k_{3} > \lambda > 1$.
The explicit controller leads 
\begin{align*}
    \delta_e&=\frac{1}{\Theta_{{L}_t}^2} \arcsin \big( \hat{\Phi}^T \psi \big) -\frac{x_t q}{V}-\alpha\label{delta_e}
\end{align*}
where $\psi$ has been defined previously, and the adaptive dynamics were presented in \ref{dPhi}. Finally, the parameters to tune are summed up
\begin{align*}
k_1< \frac{\Theta^{\alpha}+2\alpha_R\Theta^{\alpha^2}+k\Theta^{\alpha k}}{e_1\Theta^{\alpha^2}}-1\\
k_3> \frac{c_1}{\beta_{2w}V^2\cos{\alpha} \mu x_t} \quad \quad
c_1, c_4, k_4, \Gamma_\gamma> 0
\end{align*}

\noindent{\bf Parameters projection.} Another added control difficulty that we have encountered in the perching maneouvre of this tail-steered ornithopter is the tail stall. 
This corresponds to that resulting control for path angle \eqref{u} is only well defined for $\arg \arcsin \leq 1$.
To deal with that, a smooth projector--with boundary layer transition--is proposed, that forces the adaptive parameters to stay in a parameters space away from such critical condition. Let define the parameters convex space as
$
    \Pi := \{ \hat{\Phi} \in \mathbb{R}^6: \ P(\hat{\Phi})\leq \varepsilon \},
$
where $P:=(\hat{\Phi}^T \Psi )^2-(\frac{S_t}{S}-\varepsilon)^2$, which has been verified with realistic simulations that is convex in $\cal{F}$. Note that $\varepsilon \in [0,\frac{S_t}{S}]$ is the thickness of the boundary layer, defined as $\Pi_\epsilon$. It is also assumed that the boundary $\partial \Pi$ is smooth and 
from \cite{projector} a projector operator that satisfies those requirements becomes 
\begin{empheq}[left=\text{Proj($\dot{\hat\Phi}$):}\empheqlbrace]{alignat*=4}
&\dot{\hat{\Phi}}, \ & \text{if} & \ \hat{\Phi} \in \Pi \ & \text{or} & \ \nabla P^T\dot{\hat{\Phi}}\leq 0,\\
&(I_{6}-\varsigma(\hat{\Phi})\overrightarrow{\mathbf{p}})\dot{\hat{\Phi}}, \ & \text{if} & \ \hat{\Phi} \in \Pi_\varepsilon / \Pi \ & \text{and} & \ \nabla P^T\dot{\hat{\Phi}}>0,
\end{empheq}
with $\Gamma$ a positive-definite matrix, $\overrightarrow{\mathbf{p}}:=\Gamma \nabla P \nabla P^{T}/|\nabla P|_{\Gamma}^2$ the unitary vector normal to the hyperplane tangent to $\partial \Pi$ and $\varsigma(\hat{\Phi})=\min\{1,P(\hat{\Phi})/\varepsilon\}$ the weighting function that smoothes the transition in the boundary layer.

\begin{prop} \label{pr:att}
Consider the attitude error dynamics \eqref{eqe1}--\eqref{eqe3} and $\gamma_{R}(t), \dot \gamma_{R}(t), \ddot \gamma_{R}(t)\in \cal{L}_\infty$. For any control gain $k_1\in \Omega_1$ and $c_1$, $k_3$ and $\Gamma_\gamma$ positive, the adaptive state feedback \eqref{u}--\eqref{dPhi} guarantees the boundedness of $e_i$, $i=1,2,3$, and ${\hat{\Phi}}$, and the convergence of $e_i$, i.e. $\lim_{t\to \infty} |e_i(t)| = 0$.
\end{prop}
\begin{proof}
Let $e_a:=(e_1,\tilde e_2,\tilde e_3,\tilde\Phi)^T\in \mathbb{R}^9$ be the state vector. The function $W_3$ of \eqref{W3} is bounded by $\underline{\nu}(e_a) \leq W \leq \overline{\nu}(e_a)$ for some $\underline{\nu},\overline{\nu} \in \mathcal{K}_{\infty}$, and hence it is positive definite radially unbounded and from the derivations above $\dot W\leq 0$ from \eqref{dW3b}. Therefore, $e_a \in {\cal{L}}_\infty$ and by Property~\ref{P}, i.e. property of $\eta$ with $k_1\in \Omega_1$, so do $e_1,e_2,e_3,\eta, \hat\Phi \in {\cal{L}}_\infty$, under any $\gamma_{R}, \dot \gamma_{R}\in \cal{L}_\infty$.
To prove convergence first note that the closed loop \eqref{eqe1}--\eqref{eqe3} is non autonomous because of $\dot \gamma_{R}$. Thus, on the one hand, the global boundedness of $e_i, \eta$ and $\ddot \gamma_{R}$ imply that $\ddot W_3 \in \cal{L}_{\infty}$ and therefore we conclude that $e_i$ and $\eta$ are uniformly continuous. Finally, since $W_3$ is bounded from below, Barbalat's lemma guarantees that  $\lim_{t \to \infty}\left|\begin{array}{c}e_i(t) \\ \eta(t) \end{array}\right| = 0,$ $ t\geq 0$, $i=1,2,3$.
\end{proof}

\begin{remark}
We highlight here a major difference between the control of ornithopters and fixed-wing aircraft in \cite{gavilan2015}. The velocity and flight-path angle designed for the optimal perching trajectory are the references for the velocity and path angle controllers of Proposition \ref{pr:vel} and \ref{pr:att}, i.e. $V_R$ and $\gamma_{R}$, which are bounded along with their derivatives by construction. 
In \cite{gavilan2015} for fixed-wing aircraft only regulation was required, i.e. constant $V_R$ and $\gamma_{R}$, that allowed us to conclude asymptotic stability. However, in the case of ornithopters, constant $V_R$ and $\gamma_{R}$ are not feasible perching trajectories and, their dynamics are inherently non autonomous, unlike the autonomous fixed-wing aircraft there. Therefore, this work generalizes the controller of \cite{gavilan2015} to follow time-dependent references and for a flapping-wing ornithopter. From a theoretical point of view, we prove uniform stability along the entire perching trajectory, a strong and very practical property, such as the ability to withstand constant perturbations (e.g. wind gust). 
\end{remark}

\subsection{Guidance law: $\Sigma_{p}$}
The control design is complete, and it consists of the the optimal trajectory as the reference for the controllers of Proposition \ref{pr:vel} and \ref{pr:att}. However, a guidance law is needed to link the references with the inertial-frame position, thus correcting drift from kinematics \eqref{eom1}--\eqref{eom2}. 
To this end, let us define the vertical distance (altitude) error $e_G=z-z_R$, with $z_R$ as the reference. Then the error dynamics from \eqref{eom2} read $\dot{e}_G=V\sin{\gamma_{G}}-V_R \sin{\gamma_R}$, where $\gamma_{G}$ denotes the $\gamma$ of the guidance. Thus, defining the Lyapunov function $W_G:=\frac{1}{2} e_G^2$ and enforcing $\dot W_{G}<0$, it yields a $\gamma_R$ correction given by
\begin{equation*}
\gamma_G(z,V):=\arcsin{\left((V_{R}\sin{\gamma_R}-k_G e_G)/V\right)}, 
\end{equation*}
where $k_G>0$, such that $\gamma_G(z_R,V_R)\equiv\gamma_R$. Noting that $V>0$ and hence $\dot x>0$, the course coordinate is taken into account indirectly by combining \eqref{eom1} and \eqref{eom2} to obtain $e_G(x)=x \tan \gamma-x_R\tan \gamma_R$.

\section{Results}\label{sec:Results}
Regarding implementation, the closed-loop dynamics of the FW-UAV and the optimal  perching path generator have been simulated numerically using a simple Euler scheme. It is important to note that the calculation of the optimal perching path could be substituted by a look-up table if the onboard processor capability is limited, while the controller becomes just a function of states easily implementable in real time. 
\begin{figure*}[htb]
    \centering
    \includegraphics[width=0.9\textwidth]{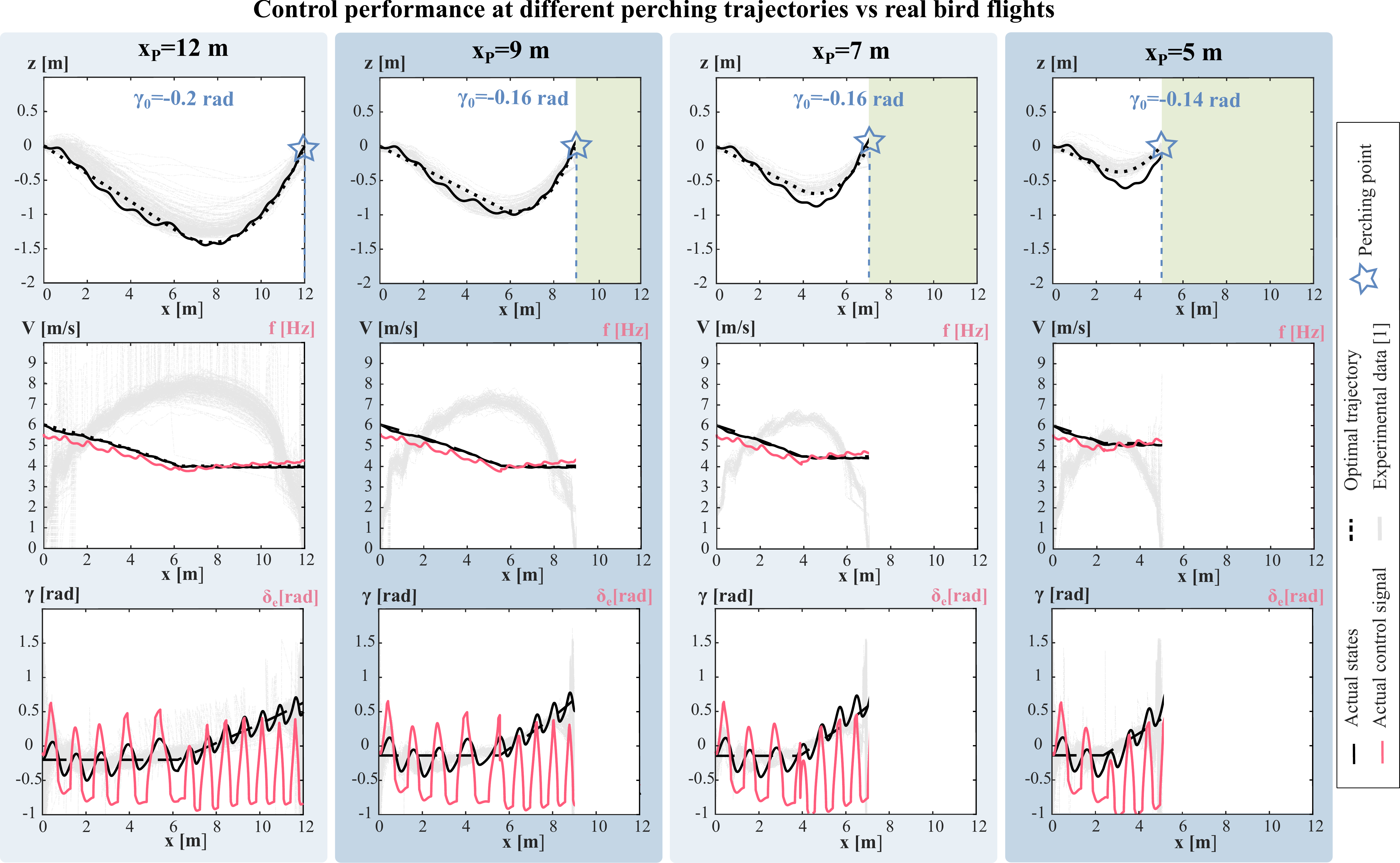}
    \caption{Control performance and comparison between proposed and real trajectories. Each column corresponds to a different simulation, with different perching distance. The initial conditions of the trajectories have been extracted from real flights. The simulated controlled paths states (black lines) and control signals (red lines) are shown. The optimal proposed trajectory (dotted line) and the Drogon hawk perching flight real dataset (grey lines) are also shown.}
    \label{fig:results}
\end{figure*}
The controller parameters have been tuned by trial and error in a few iterations, taking into account the stability limits  
and are collected in Table \ref{aero_params}.
In order to verify that minimal velocity is a bioinspired objective function, we compare natural perching flights recorded by motion capture in \cite{hawk_perching} with those optimal manoeuvres proposed in this study by optimizing perch velocity. 
We calculated the optimal path with initial conditions $\gamma_0=\{-0.2,-0.14\}$, $V_0=6$, $x_P=\{5,7,9,12\}$, $z_P=0$ estimated from the ``Drogon'' hawk perching-flight real dataset, which have similar characteristics to our platform, ${m^{E-flap}}/{m^{Drogon}}=0.97$, ${b^{E-flap}}/{b^{Drogon}}=1.47$ and ${S^{E-flap}}/{S^{Drogon}}=0.5$.

Note that while the mass is almost the same, the real bird wing's surface is much smaller. However, we assume forces involved in flight will be similar due to the high efficiency of natural flight in contrast to recently developed ornithopters.
The comparison and analysis are presented further in the results section along with Fig. \ref{fig:results}, where we analyze the proposed control performance over those optimal trajectories, to avoid duplicating data here.
With this setup four different initial conditions have been simulated. Those has been extracted from the experimental data recorded from Drogon's Hawk in \cite{hawk_perching}, as mentioned before. The results are shown in Fig. \ref{fig:results}, where we plot the path (top), velocity (middle) and path angle (bottom) of the actual simulated flight (solid black line) and reference optimal trajectory (dotted black lines). The control signals are also shown in red. Note that both the reference, actual and experimental paths are similar, increasing the error the closer the starting point is to the perching, since there is less time to maneuver. This shows both that the controller is effective in following the trajectory, and that the minimum velocity trajectory is similar to that used by birds in nature. 
On the other hand, the trajectory angle follows a similar trend, with a high pitch up in the second phase of flight. However, while the velocities range is similar, the tendency in velocity has great differences with the birds, since in the experiment the birds started from a resting position and had to make a rapid acceleration, that is, takeoff was coupled with perching. But once birds reach the velocity, about the second stage, they became similar.
\begin{figure}[ht]
    \centering
    \includegraphics[width=0.95\columnwidth]{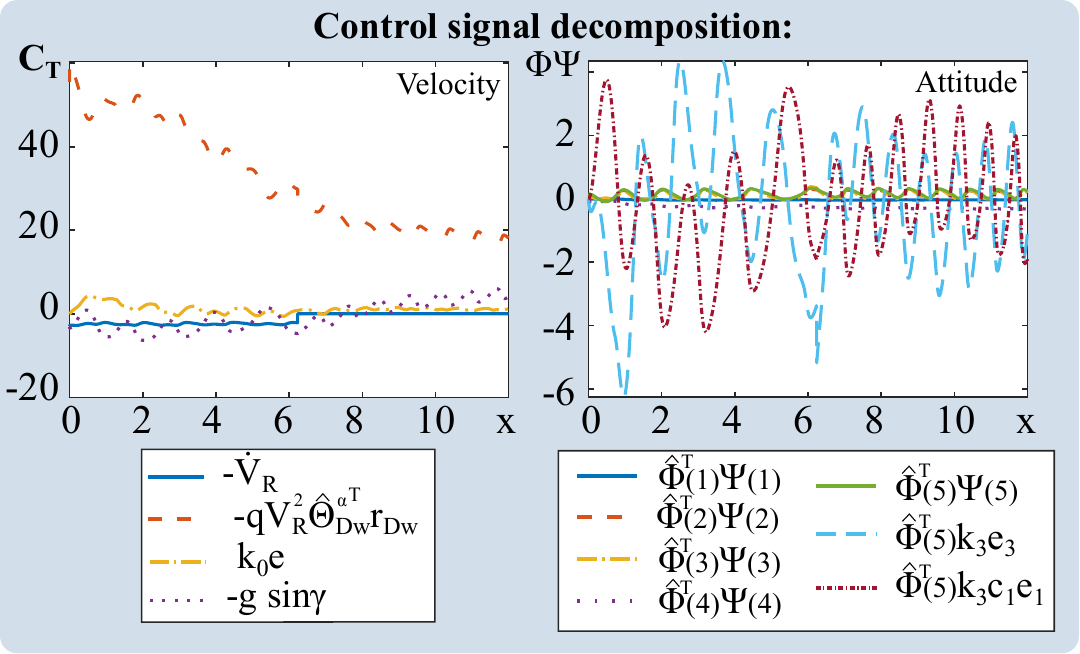} 
    \caption{Control decomposition (velocity in left, attitude in right) of flight 1 in Fig. \ref{fig:results}. }
    \label{fig:adaptive}
\end{figure}
The controller performs well with no high oscillation, in contrast to tail deflection, which is highly affected by flapping oscillations, which cannot be removed because it is the propulsion mechanism. 

The decomposition of the control signal is shown in Fig. \ref{fig:adaptive}. At the left, the velocity control decomposition is shown. As it can be seen, the adaptive contribution (dashed red) is the highest, followed by the proportional (dash-dot yellow). However, regarding attitude control (right figure), the main contributions are by the proportionals (dashed-light blue and small dashed-dot red), while the adaptive remains smaller. Note that these relative contributions depend on the control tune. However, we could deduce that manual tuning has led to the adaptive being less agressive and effective in the attitude controller for control due to the large oscillations in flapping. Note that in Fig. \ref{fig:adaptive} the fluctuations due to flapping are smaller in velocity control than in attitude.

Control is robust against perturbations, since flapping induces considerable periodic oscillations. 
In addition, robustness tests have been carried out on changes in the initial guess of the adaptive parameters, changes in the weight of the aircraft and center of gravity, providing satisfactory results even with 100\% change. The results are not presented here due to the space limitation.

\section{Conclusions and future work}\label{Conclusions and future work}
An analytical optimal maneuver has been developed for the approach and perch trajectories of the ornithopter, in conjunction with a nonlinear adaptive controller to perform safely and efficiently. Realistic simulations demonstrate the proposed scheme ensures feasibility and good performances of control, with a perching error less than 10 cm. By comparing to real Hawk perching from \cite{hawk_perching}, trajectories are similar, highlighting that the minimal end velocity trajectory is a bioinspired objective through perching. This lays the foundation for reliable onboard controllers in FW-UAVs and provides a step forward to fully understanding these agile bird maneuvers.

The optimal trajectory proposed is independent of the dynamic model, which is convenient due to inaccuracies in modeling. It just depends on initial conditions and dynamic and kinematic limitations of the platform. Thus, given the optimal deceleration ratio and turn velocity as a function of initial conditions (for example in a look-up table), it can be calculated on board the optimal path for the control reference. Moreover, the perching analysis proposed here could help UAV designers select the dynamic limitations of the platforms.

Regarding future work, it would be necessary to increase the bird data recorded throught perching, including wider range of initial conditions and longer flight to decouple take-off and perching, which would shed more light on bioinspired perching flights. However, researchers point out that the birds perform a forward sweep to make a quick pitch up \cite{mechanics_perching}, in addition to gaining control with the extension of the tail. Clearly, efficient lightweight implementations in FW-UAVs of morphing in wings and tails could increase maneuverability in perching, by augmenting the control surfaces. 

\bibliographystyle{IEEEtran}
\bibliography{biblio_final.bib}

\begin{thebibliography}{10}
\providecommand{\url}[1]{#1}
\csname url@samestyle\endcsname
\providecommand{\newblock}{\relax}
\providecommand{\bibinfo}[2]{#2}
\providecommand{\BIBentrySTDinterwordspacing}{\spaceskip=0pt\relax}
\providecommand{\BIBentryALTinterwordstretchfactor}{4}
\providecommand{\BIBentryALTinterwordspacing}{\spaceskip=\fontdimen2\font plus
\BIBentryALTinterwordstretchfactor\fontdimen3\font minus
  \fontdimen4\font\relax}
\providecommand{\BIBforeignlanguage}[2]{{%
\expandafter\ifx\csname l@#1\endcsname\relax
\typeout{** WARNING: IEEEtran.bst: No hyphenation pattern has been}%
\typeout{** loaded for the language `#1'. Using the pattern for}%
\typeout{** the default language instead.}%
\else
\language=\csname l@#1\endcsname
\fi
#2}}
\providecommand{\BIBdecl}{\relax}
\BIBdecl

\bibitem{hawk_perching}
M.~KleinHeerenbrink, L.~A. France, C.~H. Brighton, and G.~K. Taylor,
  ``Optimization of avian perching manoeuvres,'' \emph{Nature}, no. 7917, 2022.

\bibitem{mechanics_perching}
A.~C. Carruthers, A.~L.~R. Thomas, S.~M. Walker, and G.~K. Taylor, ``Mechanics
  and aerodynamics of perching manoeuvres in a large bird of prey,'' \emph{The
  Aeronautical Journal}, no. 1161, pp. 673--680, 2010.

\bibitem{bio_informed}
C.~Harvey, G.~Croon, G.~Taylor, and R.~Bomphrey, ``Lessons from natural flight
  for aviation: then, now and tomorrow,'' \emph{The Journal of experimental
  biology}, vol. 226, 04 2023.

\bibitem{harvey_morfing}
C.~Harvey, L.~L. Gamble, C.~R. Bolander, D.~F. Hunsaker, J.~J. Joo, and D.~J.
  Inman, ``A review of avian-inspired morphing for uav flight control,''
  \emph{Progress in Aerospace Sciences}, vol. 132, p. 100825, 2022.

\bibitem{Ruiz2022}
C.~Ruiz, J.~{\'A}. Acosta, and A.~Ollero, ``Optimal elastic-wing for
  flapping-wing robots through passive morphing,'' \emph{Robotics and
  automation letters}, 2022.

\bibitem{optimal_kinematics}
B.~Parslew, ``Predicting power-optimal kinematics of avian wings,'' \emph{J. R.
  Soc. Interface}, 2015.

\bibitem{optperchfixed}
A.~Wickenheiser and E.~Garcia, ``Optimization of perching maneuvers through
  vehicle morphing,'' \emph{Journal of Guidance Control and Dynamics}, vol.~31,
  pp. 815--823, 07 2008.

\bibitem{perching_reinforcement}
A.~Waldock, C.~Greatwood, F.~Salama, and T.~Richardson, ``Learning to perform a
  perched landing on the ground using deep reinforcement learning,''
  \emph{Journal of Intelligent \& Robotic Systems}, vol.~92, 12 2018.

\bibitem{Feliu2021}
D.~Feliu-Talegon, J.~{\'A}. Acosta, and A.~Ollero, ``Control aware of
  limitations of manipulators with claw for aerial robots imitating bird's
  skeleton,'' \emph{IEEE Robotics and Automation Letters}, vol.~6, no.~4, pp.
  6426--6433, 2021.

\bibitem{khalil}
H.~K. Khalil, \emph{Nonlinear Systems}.\hskip 1em plus 0.5em minus 0.4em\relax
  Prentice Hall, 2002.

\bibitem{gavilan2015}
F.~Gavilan, R.~Vazquez, and J.~A. Acosta, ``Adaptive control for aircraft
  longitudinal dynamics with thrust saturation,'' \emph{Journal of Guidance,
  Control, and Dynamics}, vol.~38, no.~4, pp. 651--661, 2015.

\bibitem{maldonado2020}
F.~J. Maldonado, J.~{\'A}. Acosta, J.~Tormo-Barbero, M.~Guzman, and A.~Ollero,
  ``Adaptive nonlinear control for perching of a bioinspired ornithopter,''
  \emph{2020 IEEE International Conference on Intelligent Robots and Systems},
  2020.

\bibitem{adaptive_insect}
A.~Banazadeh and N.~Taymourtash, ``Adaptive attitude and position control of an
  insect-like flapping wing air vehicle,'' \emph{Nonlinear Dynamics}, vol.~85,
  pp. 47--66, 2016.

\bibitem{weiNN}
W.~He, Z.~Yan, C.~Sun, and Y.~Chen, ``Adaptive neural network control of a
  flapping wing micro aerial vehicle with disturbance observer,'' \emph{IEEE
  Transactions on Cybernetics}, vol.~47, no.~10, pp. 3452--3465, 2017.

\bibitem{chenNN}
C.~Qian, Y.~Fang, and Y.~Li, ``Neural network-based hybrid three-dimensional
  position control for a flapping wing aerial vehicle,'' \emph{IEEE
  Transactions on Cybernetics}, vol.~53, no.~10, pp. 6095--6108, 2023.

\bibitem{weiADAPT}
H.~Gao, W.~He, Y.~Zhang, and C.~Sun, ``Adaptive finite-time fault-tolerant
  control for uncertain flexible flapping wings based on rigid finite element
  method,'' \emph{IEEE Transactions on Cybernetics}, no.~9, 2022.

\bibitem{weiIT}
W.~He, T.~Meng, X.~He, and C.~Sun, ``Iterative learning control for a flapping
  wing micro aerial vehicle under distributed disturbances,'' \emph{IEEE
  Transactions on Cybernetics}, vol.~49, no.~4, pp. 1524--1535, 2019.

\bibitem{perching}
R.~Zufferey, J.~Tormo-Barbero, D.~Feliu-Taleg\'{o}n, S.~Nekoo, J.~A. Acosta,
  and A.~Ollero, ``How ornithopters can perch autonomously on a branch,''
  \emph{Nature communications}, vol.~13, no. 7713, 2022.

\bibitem{Ruiz2021}
C.~Ruiz, J.~{\'A}. Acosta, and A.~Ollero, ``Aerodynamic reduced-order volterra
  model of an ornithopter under high-amplitude flapping,'' \emph{Aerospace
  Science and Technology Journal}, vol. 121, p. 107331, 2021.

\bibitem{guzman2021}
M.~Guzm{\'a}n, C.~Ruiz, F.~J. Maldonado, R.~Zufferey, J.~Tormo-Barbero,
  J.~{\'A}. Acosta, and A.~Ollero, ``Design and comparison of tails for
  bird-scale flapping-wing robots,'' in \emph{2021 IEEE/RSJ International
  Conference on Intelligent Robots and Systems (IROS)}, 2021, pp. 6358--6365.

\bibitem{Zufferey2021}
R.~Zufferey, J.~Tormo-Barbero, M.~del Mar~Guzm{\'a}n, F.~J. Maldonado,
  E.~Sanchez-Laulhe, P.~Grau, M.~P{\'e}rez, J.~{\'A}. Acosta, and A.~Ollero,
  ``Design of the high-payload flapping wing robot e-flap,'' \emph{IEEE
  Robotics and Automation Letters}, vol.~6, pp. 3097--3104, 2021.

\bibitem{carlos_aero}
C.~de~Cos and J.~{\'A}. Acosta, ``Explicit aerodynamic model characterization
  of a multirotor unmanned aerial vehicle in quasi-steady flight,''
  \emph{Journal of Computational and Nonlinear Dynamics}, vol.~15, 05 2020.

\bibitem{Lanchester}
F.~Lanchester, \emph{Aerodonetics}, ser. A Complete Work on Aerial
  Flight.\hskip 1em plus 0.5em minus 0.4em\relax London: Constable, 1908.

\bibitem{TAYLORTHOMAS2002}
G.~Taylor and A.~Thomas, ``Animal flight dynamics ii. longitudinal stability in
  flapping flight,'' \emph{Journal of Theoretical Biology}, no.~3, 2002.

\bibitem{projector}
M.~Krsti{\'c} and P.~Kokotovi{\'c}, \emph{Nonlinear and Adaptive Control
  Design}.\hskip 1em plus 0.5em minus 0.4em\relax New York, NY:
  Wiley-Interscience, 1995.

\end{thebibliography}

\begin{appendices}
\section{Aerodynamic Model Parameters}\label{ap:aerodynamics}
The parameters values of the E-flap FW-UAV  model used in Section \ref{sec:dynmodel} are presented in Table \ref{aero_params}. 

\begin{table*}[htb]
\begin{tabular}{p{-1cm}p{-1cm}p{-1cm}p{-1cm}p{-1cm}p{-1cm}l|ll|}
\cline{1-6} \cline{8-9}
\multicolumn{6}{|l|}{Geometric and mass parameters} &  & \multicolumn{2}{l|}{Aerodynamic parameters} \\ \cline{1-6} \cline{8-9} 
\multicolumn{1}{|l}{$x_a$} & \multicolumn{1}{l|}{$0.05$} & $I_y$ & \multicolumn{1}{l|}{$0.044$} & $S$ & \multicolumn{1}{l|}{$0.42$} &  & $\Theta_{{L}_w}$ & $[-0.31 \hspace{0.2cm}2.19\hspace{0.2cm} 0.37 \hspace{0.2cm}7.38 \hspace{0.2cm}-0.37\hspace{0.2cm} 2.33\hspace{0.2cm} 0]$ \\ \cline{1-6} \cline{8-9} 
\multicolumn{1}{|l}{$x_t$} & \multicolumn{1}{l|}{$0.3$} & $g$ & \multicolumn{1}{l|}{$9.81$} & $m$ & \multicolumn{1}{l|}{$0.64$} &  & $\Theta_{\check{L}_w}$ & $[2.85\hspace{0.2cm} 2.65\hspace{0.2cm} -4.32\hspace{0.2cm} -11.24 \hspace{0.2cm}7.47 \hspace{0.2cm}-1.22 \hspace{0.2cm}0]$ \\ \cline{1-6} \cline{8-9} 
\multicolumn{1}{|l}{$c $} & \multicolumn{1}{l|}{$0.36$} & $\rho$ & \multicolumn{1}{l|}{$1.22$} & $S_t$ & \multicolumn{1}{l|}{$0.12$} &  & $\Theta_{{D}_w}$ & $[4.41\hspace{0.2cm} -1.10\hspace{0.2cm} -16.47\hspace{0.2cm}  7.21\hspace{0.2cm} 24.82\hspace{0.2cm} 0.76\hspace{0.2cm} -12.44]$ \\ \cline{1-6} \cline{8-9} 
 &  &  &  &  &  &  & $\Theta_{{M}_w}$ & $[-6.14\hspace{0.2cm} -12\hspace{0.2cm} 6.97\hspace{0.2cm} -41\hspace{0.2cm} -7.97\hspace{0.2cm} -74]\cdot10^{-2}$ \\ \cline{1-6} \cline{8-9} 
\multicolumn{6}{|l|}{Control parameters} &  & $[\Theta_{{L}_t}^1 \hspace{0.2cm} \Theta_{{L}_t}^2]$ & $[0.94 \hspace{0.2cm}2.92]$ \\ \cline{1-6} \cline{8-9} 
\multicolumn{1}{|l}{$k_0$} & \multicolumn{1}{l|}{4} & $\Gamma_V  $ & \multicolumn{1}{l|}{. [0.5 0.2]$I_2$} & $k_G$ & \multicolumn{1}{l|}{4} &  & $[\Theta_{{D}_t}^0 \hspace{0.2cm} \Theta_{{D}_t}^1 \hspace{0.2cm} \Theta_{{D}_t}^2]$ & $[0.36\hspace{0.2cm} 0.32\hspace{0.2cm} 4.23]$ \\ \cline{1-6} \cline{8-9} 
\multicolumn{1}{|l}{$c_1$} & \multicolumn{1}{l|}{3} & $k_3$ & \multicolumn{1}{l|}{0.5} & $K_1$ & \multicolumn{1}{l|}{0.5} &  & $[\Theta_{{M}_t}^1 \hspace{0.2cm} \Theta_{{M}_t}^2]$ & $[-0.65\hspace{0.2cm} 2.26]$ \\ \cline{1-6} \cline{8-9} 
\multicolumn{1}{|l}{$\Gamma_\gamma$} & \multicolumn{1}{l|}{0.05 $I_6$} & $\varepsilon$ & \multicolumn{1}{l|}{0.2} & $\Gamma$ & \multicolumn{1}{l|}{0.01 $I_6$} &  & $\Theta_{{L}_w}^\prime$ & $[-9.88\cdot10^{-2} \hspace{0.2cm}2.14\hspace{0.2cm} 7.39 \hspace{0.2cm}-0.24\hspace{0.2cm} 2.37]$ \\ \cline{1-6} \cline{8-9} 
 &  &  &  &  &  &  & $\Theta_{{D}_w}^\prime$ & $[2.17\hspace{0.2cm} 7.09 \hspace{0.2cm}-1.92]$ \\ \cline{8-9} 
 &  &  &  &  &  &  & $[\Theta_{S}^1 \hspace{0.2cm} \Theta_{S}^2]$ & $[1.35\hspace{0.2cm} -0.3]$ \\ \cline{8-9} 
\end{tabular}
\caption{Dynamic model and control parameters}
\label{aero_params}
\end{table*}

\section{Optimal trajectory solution}\label{ap:solutions}
The minimal velocity perching problem is described by the following system of 16 equations and unknowns ($\dot{V}_D, \dot{\gamma}_T, \gamma_P, V_P$, $\sigma_{1-5}$  $\lambda_{1-7})$:
\begin{align*}
 \lambda_1+\lambda_2 \tan{\gamma_0} -  \lambda_3 \left( \frac{2 \dot{V}_D^2}{\cos{\gamma_0}(V_P^2-V_0^2)} \right)&=0,\\
 \lambda_1+\lambda_2 \tan{\gamma_M}+\lambda_4 \left( \frac{\dot{\gamma}_T^2}{V_P(\sin{\gamma_P}-\sin{\gamma_0})} \right)&=0,\\
  -\lambda_1 \left(\frac{V_P}{\dot{\gamma}_T}\cos{\gamma_P}\right)-\lambda_2 \left(\frac{V_P}{\dot{\gamma}_T}\sin{\gamma_P}\right) - \lambda_5+\lambda_6&=0,\\
\dot{V}_D \dot{\gamma}_T - \lambda_1 \left(\dot{V}_D(\sin{\gamma_P}+\sin{\gamma_0})-\cos{\gamma_0} V_P \dot{\gamma}_T \right) & \\
 + \ \lambda_2 \left(\dot{V}_D(\cos{\gamma_P}-\cos{\gamma_0})-\sin{\gamma_0} V_P \dot{\gamma}_T \right) 
 - \lambda_7 \dot{V}_D \dot{\gamma}_T  &=0,\\
 2\lambda_3\sigma_1= 2\lambda_4\sigma_2=2\lambda_5\sigma_3=2\lambda_6\sigma_4=
 2\lambda_7\sigma_5&=0,\\
 x_P-\frac{V_P}{\dot{\gamma}_T}(\sin{\gamma_P}-\sin{\gamma_0})-\frac{\cos{\gamma_0}}{2{\dot{V}_D}}(V_P^2-V_0^2)&=0,\\
 z_P+\frac{V_P}{\dot{\gamma}_T}(\cos{\gamma_P}-\cos{\gamma_0})-\frac{\sin{\gamma_0}}{2{\dot{V}_D}}(V_P^2-V_0^2)&=0,\\
 \dot{V}_D^{min}-\dot{V}_D+\sigma_1^2=0, \quad
 \dot{\gamma}_T-\dot{\gamma}_T^{max}+\sigma_2^2&=0, \\
 \gamma_P^{min}-\gamma_P+\sigma_3^2=0, \quad
 \gamma_P-\gamma_P^{max}+\sigma_4^2&=0, \\
 V_P^{min}-V_P+\sigma_5^2&=0.
\end{align*}
After analyzing the $\lambda$ and $\sigma$ equations, six possible solutions are obtained. The analytical solution for each case is presented in detail in Section \ref{Optimal perching maneuvers}.

\end{appendices}

\end{document}